\documentclass[12pt,preprint]{aastex}

\usepackage{natbib}
\usepackage{multirow}
\usepackage{pdfpages}
\usepackage{graphicx}
\usepackage{fancyvrb}
\usepackage{epstopdf}
\usepackage[mathscr]{euscript}
\usepackage{framed}
\usepackage{amsmath}
\usepackage{float}

\slugcomment{}

\shorttitle{}
\shortauthors{Brasseur et al.} 

\VerbatimFootnotes

\makeatletter
\g@addto@macro\@floatboxreset\centering
\makeatother

\begin{document}
\title{Short Duration Stellar Flares in GALEX Data}
\shorttitle{GALEX Flares}
 
\author{C. E. Brasseur}
\affil{Space Telescope Science Institute}
\email{cbrasseur@stsci.edu}
\author{Rachel A. Osten\altaffilmark{1}}
\affil{Space Telescope Science Institute}
\email{osten@stsci.edu}
\altaffiltext{1}{Also at Center for Astrophysical Sciences, Johns Hopkins University, Baltimore, MD 21218}
\author{Scott W. Fleming}
\affil{Space Telescope Science Institute}
\email{fleming@stsci.edu}

\begin{abstract}
  We report on a population of short duration near-ultraviolet (NUV) flares in stars observed by the Kepler and GALEX missions.  We analyzed NUV light curves of 34,276 stars observed from 2009-2013 by both the GALEX (NUV) and Kepler (optical) space missions with the eventual goal of investigating multi-wavelength flares. From the GALEX data we constructed light curves with a 10 second cadence, and ultimately detected 1,904 short duration flares on 1,021 stars. The vast majority (94.5\%) of these flares have durations less than five minutes, with flare flux enhancements above the quiescent flux level ranging from 1.5 to 1700. The flaring stars are primarily solar-like, with T$_{\rm eff}$ ranging from 3,000-11,000 K and radii between 0.5-15 R$_{\odot}$. This set of flaring stars is almost entirely distinct from that of previous flare surveys of Kepler data and indicates a previously undetected collection of small flares contained within the Kepler sample. The range in flare energies spans 1.8$\times$10$^{32}$-8.9$\times$10$^{37}$ erg, with associated relative errors spanning 2-87\%. The flare frequency distribution by energy follows a power-law with index $\alpha=1.72\pm0.05$, consistent with results of other solar and stellar flare studies at a range of wavelengths. This supports the idea that the NUV flares we observe are governed by the same physical processes present in solar and optical flares. The relationship between flare duration and associated flare energy extends results found for solar and stellar white-light flares, and suggests that these flares originate in regions with magnetic field strengths of several hundred Gauss, and length
  scales of order 10$^{10}$ cm.
  
\end{abstract}


\section{Introduction }
Magnetic activity on Sun-like stars has many observational manifestations, ranging from evidence of starspots and magnetic fields, to emissions associated with non-radiative heating of the stellar atmosphere, to particle acceleration \citep{donatilandstreet2009}. A reconfiguration of the magnetic field during a magnetic reconnection event causes liberation of energy \citep{benzgudel2010} which goes into plasma heating, particle acceleration, shocks, and mass motions. The transient brightening in emission is the most obvious marker of a flare. Flares are the most dramatic energy release events that cool stars will experience while on the main sequence. Stellar flares are one manifestation of magnetic activity which is demonstrated by all solar-like stars to varying degrees. This has led to solar-like flares being detected in cool stars with outer convection zones over a range of stellar ages, from X-ray flares on $\sim$1 MY old solar-like stars \citep{wolketal2005} to optical flares on stars several GY older than the Sun \citep{paulson2006,ostenetal2012}, and across a range of stellar masses, from at and below the hydrogen-burning limit \citep{stelzer2006,gizis2017} to stars with thin outer convection zones \citep{balona2015}.

A stellar flare originates from the reconfiguration of the magnetic field in the outer atmosphere. Because all layers of the stellar atmosphere are thought to be involved in a flare, and several physical processes are involved, emissions across the electromagnetic spectrum can be produced.   X-ray flares can increase by orders of magnitude above the underlying coronal emission \citep{osten2016}, and last from tens of minutes to hours.  Visible light flares can also produce dramatic increases \citep{maehara2012}, and blue-optical flares on M dwarfs have long been known to produce increases of up to factors of thousands on short time-scales \citep{kowalski2010}. Radio flares originating from incoherent gyrosynchrotron emission \citep{osten2005} and coherent emission \citep{ostenbastian2006} indicate the presence and action of accelerated particles in stellar flares. Study of the ultraviolet (UV) spectral region has been available for stellar flares since the days of the  International Ultraviolet Explorer (IUE) with a low temporal cadence e.g. \citet{haisch1981}. UV instruments on Hubble continued the exploration, revealing  flares with different spectroscopic signatures in the FUV seen at high cadence \citep{gdec2002}. Early results from the Galaxy Evolution Explorer (GALEX) utilized high time resolution observations to reveal evidence for  flare oscillations \citep{welsh2006}.

The ultraviolet spectral region probes chromospheric plasma with temperatures near 10$^{5}$K; solar flare emissions in this spectral region correspond to the impulsive phase of a flare, when energization mechanisms are most important, and electron beams may be responsible for energy deposition \citep{fletcher2003}.  \citet{kleint2016} explored co-spatial time-delays of a solar flare NUV continuum emission occurring after hard X-ray emission, which appear to be compatible with electron beam heating scenarios.  \citet{kowalski2017} demonstrated that for a large solar flare, the atmospheric response to a high beam flux density is able to reproduce the observed continuum brightness in the NUV, by comparing observed spectra to those calculated with a radiative hydrodynamic code. The timescales for such energy deposition can be quite small, on the order of tens of seconds, requiring high time resolution observations to study it.

Studying the NUV spectral region is important for learning about the emission processes produced during stellar flares, both specific to this wavelength region, and for understanding the interrelationship between energy release in the UV and optical wavelength regions during flares. The main component of the flare spectral energy distribution in the UV-optical wavelength range is often assumed to be dominated by a hot ($\sim$10$^{4}$K) blackbody continuum, as described by \citet{hawleypettersen}. \citet{ostenwolk2015} explored energy partitions using previously published data, and estimated the relative contribution of hot blackbody emission to coronal emission, providing a framework for determining the energy fractionation in particular bandpasses. \citet{kowalski2018} more recently showed specifically for the NUV spectral region that the assumption of only a hot blackbody contributing to the NUV flare emission could underestimate the radiated energy in the NUV region by factors of 2-3, and revealed open questions for this spectral region.

Establishing the amount of NUV flare emissions relative to optical (and ideally, bolometric) informs not only a deeper understanding of stars and their flares, but also contributes key information when extrapolating the impact of stellar flares on close-in exoplanets. 
\citet{France2013} motivated renewed interest in studying stellar UV emissions, by noting the importance of stellar NUV irradiation of exoplanet atmospheres in regulating the atmospheric chemistry and temperature profiles. 
The peak of the cross section for ozone photodissociation is at 2600 \AA, so characterizing NUV flare radiation is crucial to understanding the potential 
astrobiological impact of flares \citep{tian2014}.
Recently, \citet{AfrinBadhan2019} described how stellar UV activity can potentially photolyze stratospheric H$_{2}$O, depending on how well-mixed the planet's atmosphere is. 
Impulsive flares can cause orders of magnitude increases in the UV flux over the underlying quiescent activity values \citep[such as the factor of $\sim$1000 increase in GJ 3685A;][]{robinson2005}.  
Exoplanet transit surveys such as Kepler have discovered more than $\sim$800,000 flare events in the optical \citep{Davenport2016}, with the follow-on K2 mission set to find even more, and new TESS results will likely eclipse those numbers of flares detected. Having a way to relate observed flare energy in these optical bands with the likely UV contribution will be key to understanding the impact of these flares on the atmospheres of exoplanets.

Solar studies have revealed the utility of simultaneous multiwavelength observations to investigate the flare process \citep{dennisschwartz1989}. Stellar flares exhibit multi-wavelength correlations and energy partitions similar to those seen in solar flares, suggesting that the same physical process is occurring in both \citep{ostenwolk2015}, but detailed studies can reveal the extent of the similarities. This is necessary to understand the connection between the well-studied solar flares (reaching a maximum of about 10$^{32}$ erg), and the much more energetic events seen on other stars (flares on single stars can reach to 10$^{36}$ erg). Despite the tremendous advantage to studying flares at both wavelengths at the same time, such multi-wavelength observations are difficult to arrange, and few exist. The short timescales involved in flares, which can last minutes to hours, requires strictly simultaneous observations. The traditional approach is to pick a high flaring rate star and organize an observing campaign around it. While this is usually successful, this leads to biases in understanding only the most magnetically active stars, e.g. \citet{Hawley2003} and \citet{osten2005}. Other observing approaches take advantage of the proximity of young magnetically active stars in star-forming regions, and use a multiplexing advantage in observing simultaneously at multiple wavelengths to study flares in these objects, e.g. \citet{flaccomio2018,guarcello2019}. For these objects, the time domain interpretation is complicated because in very young stars ( $\lesssim$10 MY), time-varying emissions can result from accretion as well as magnetic activity. Time domain surveys that overlap in both time and direction on the sky provide a distinct set of advantages over the other two approaches. The number of targets increases by an order or magnitude or more compared to the multiplexing case, and there is no bias for high flaring rate stars. The total monitoring time can be quite long, depending on the exact nature of the time domain surveys.

In overlapping time-domain surveys, there are distinct advantages to using UV time-domain information in concert with optical data.  Both probe stellar flare emission associated with the sub-coronal stellar atmosphere, where energy deposition occurs. The target selection for the Kepler mission is biased toward solar-like stars, which broadens the stellar population probed with multi-wavelength flare studies.  The present paper takes advantage of new software \citep[the ``gPhoton'' package;][]{million2016}, which makes the calibrated photon events collected from GALEX available to construct images or light curves at specified time bins. In this paper, we detect and characterize UV flares seen with GALEX in a population of stars that has simultaneous optical data from the Kepler mission - a situation that arose whenever both telescopes happened to be observing in the same part of the sky. A forthcoming paper connects the UV and optical data for multi-wavelength flare study of solar-like stars, while this paper focusses on the GALEX detections. The paper is organized as follows: \S 2 describes the data reduction, \S 3 describes the data analysis, \S 4 discusses the findings and implications, and \S5 concludes. 

\section{Data Reduction, Sample Selection, and Flare Filtering }
     

The Galaxy Evolution Explorer (GALEX) was an orbiting space telescope active between April 2003 and June 2013 \citep{galexref}. GALEX was equipped with a 50 cm Ritchey-Chr\'etien telescope simultaneously feeding FUV and NUV detectors, with both imaging and slitless spectroscopy modes.  Each detector consisted of a stack of three microchannel plates (MCPs) that output time-tagged photon events with a time resolution of 5 thousandths of a second and spatial resolution of 4-6 arcseconds. The main Kepler mission ran from 2009-2013; due to the failure of GALEX's FUV camera in May 2009, all GALEX data simultaneous with Kepler observations are NUV only, spanning the range  1771-2831 \AA\ \citep{Morrissey_2007}.

\begin{figure}[!htb]
    \centering
    \includegraphics[width=\textwidth]{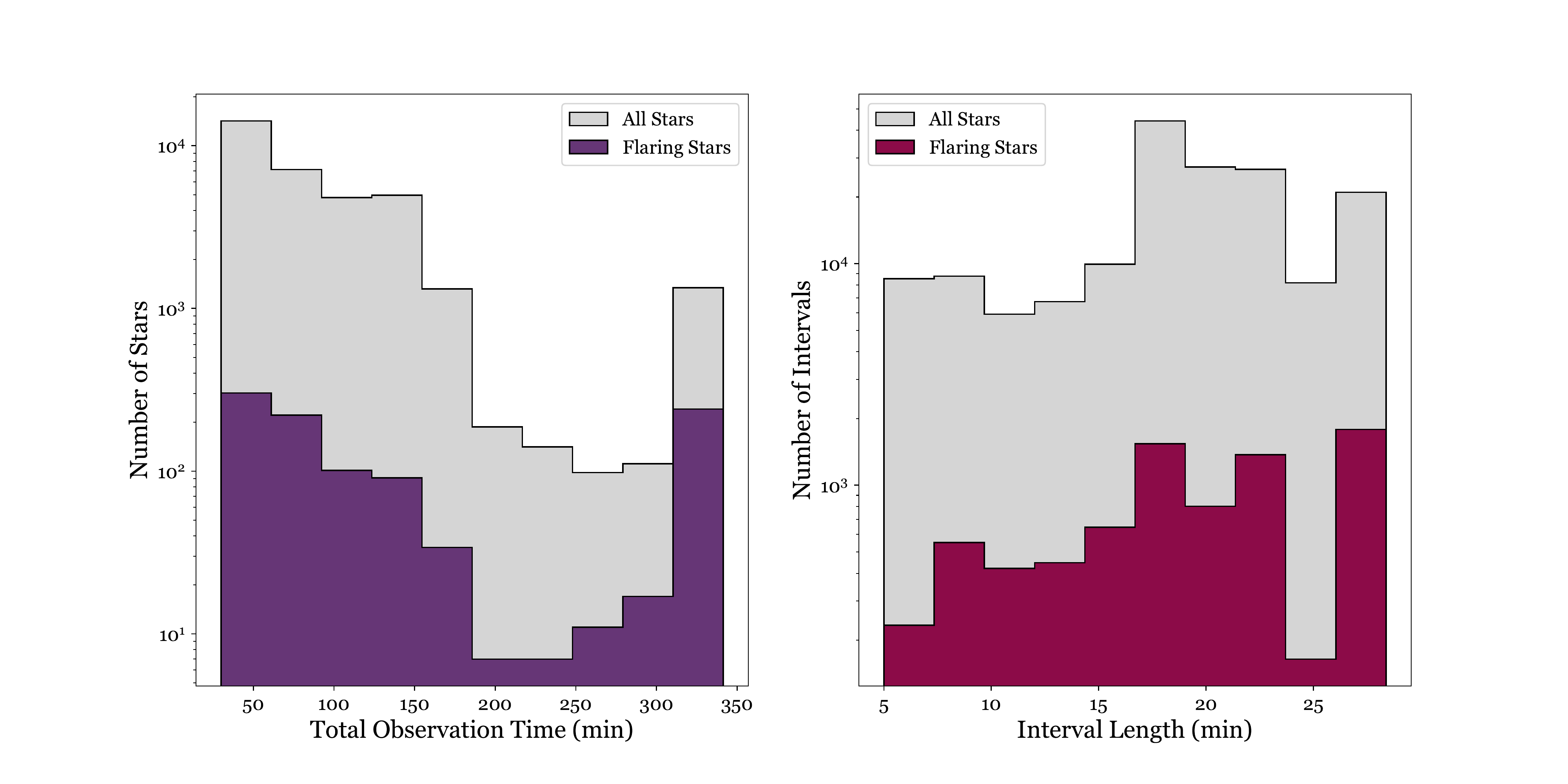}
    \caption{\textit{Left:} Histogram showing the distribution of total GALEX observation time for our sample of stars observed by both GALEX and Kepler in the 2009-2013 timeframe; the total observation time spans 30-341.5 min for this sample. Also shown is the distribution of total observation times for stars determined to contain at least one flare, after the filtering described in \S 2.1.
    \textit{Right:} Histogram showing the distribution of observation interval length, for all stars as well as the subset for which flares were identified. The maximum interval length of 30 minutes is due to the orbit of the telescope. 
    }
    \label{fig:obsHist}
\end{figure}

We compiled a list of 59,366 objects observed by both GALEX and Kepler from the Consolidated Input Values table in \citet{Huber2014}, cross-matched against the MAST KGMatch table\footnote{https://archive.stsci.edu/pub/kepler/catalogs/README\_KGMATCH}. The Consolidated Input Values table provides data about all objects targeted by Kepler, while the KGMatch table lists all objects in the Kepler Input Catalog that were targeted by GALEX. We 
extracted the observation times for each of the objects targeted by both Kepler and GALEX and considered only objects with at least 30 minutes of total GALEX observation time, consisting of continuous intervals at least 5 minutes in duration.  We also removed objects whose observation times did not fall within the Kepler mission time-range (2009-2013), and thus could not have simultaneous Kepler and GALEX observations. This left us with a final list of 34,276 targets simultaneously\footnote{We did not consider Kepler quarters as distinct, so this criteria does not guarantee truly simultaneous observations.} observed by Kepler and GALEX. Figure \ref{fig:obsHist} shows the distribution of total observation times per star and continuous interval length per number of intervals. The bulk of the sample consisted of stars with 1-2 hours of data.

The original GALEX direct imaging pipeline created integrated images from the time-tagged photon events.  More recently, a suite of tools was released called ``gPhoton,'' that allows for more flexibility in use of the GALEX dataset \citep{million2016}. Among other data products, gPhoton allows users to construct light curves at any desired cadence, allowing for time-domain studies of GALEX data such as our own. We used gPhoton to extract light curves for each of our 34,276 simultaneous  GALEX/Kepler targets. With gPhoton, light curves can be created using a variety of parameters, for this study we used 10 second binning and an aperture radius of 0.004$^{\circ}$.
We used the fields t\_mean (mean timestamp of events within the aperture, in GALEX time), exptime (estimated effective exposure time), flux (flux density within the aperture, uncorrected for background, in erg~s$^{-1}$ cm$^{-2}$ \AA$^{-1}$), flux\_err (estimated 1-sigma error in flux value, in the same units as the flux), mcat\_bg (estimated background brightness from the visit-level Merged Catalog of Sources (MCAT), scaled to the area of the aperture) \citep{million2016}. We opted to use the MCAT background estimations because they are stable over the course of a visit, and don't suffer from contamination from the source star that can happen when using an annulus, particularly as the star brightens in a flare. We also applied aperture correction, using gPhoton's built in functionality for that purpose. Our light curves were extracted using gPhoton v1.26.0, however because there have been several important updates and bug fixes since that time, we re-extracted and analyzed the light curves for our flaring stars using gPhoton v1.28.9 (the most recent version as of the time of submission), and found no appreciable difference in our results.

The steps used in filtering the initial set of target light curves to identify flares for further study are summarized in Table~\ref{tbl:dataRedux}. Step~1, described above, selects the initial pool of targets to search for flares. Step 2, the first pass automatic flare detection algorithm used the mean of the entire light curve to approximate the quiescent flux and compared it to the number of counts in each bin assuming Poisson noise on the count measurements (where the standard deviation is $\sigma = \sqrt{count}$), requiring
a maximum photon count greater than or equal to $3.5\sigma$ above the mean, and at least one adjacent point greater than or equal to $2\sigma$ above the mean.
This returned 5,591 targets with evidence of flaring. Manual examination of the light curves after this step led  to a revision in the flare detection algorithm, and the use of the median rather than the mean of the light curve as a better approximation of quiescence.
Additionally, we required that the maximum photon count be further from the median than the minimum photon count. The change in flare detection measurement from sigmas-from-mean to sigmas-from-median enabled a better approximation of the quiescent flux of the star. The new requirement that the maximum be further from the median than the minimum guards against light curves showing dips in magnitude being falsely marked as flaring objects. Application of this Step 3 to the 5,591 candidates reduced the number to 2,356 stars potentially showing evidence of flaring. 


In Step 4 of our flare filtering process, 
we  divided the light curves up into ``continuous intervals:'' intervals for which consecutive time bins are no more than 1,600 sec apart. The value of 1,600 sec was chosen because it approximates the GALEX mission bookkeeping of ``visits.''
Within each interval we considered points with counts greater than or equal to $3.5\sigma$ above the global median (calculated across the entire light curve) with at least one adjacent point greater than or equal to $2\sigma$ above the median to be flare peaks. We identified the flare edges as the first points to either side of the peak less than or equal to the global median, or the edge of the interval, whichever was reached first.  If the entire interval was identified as flaring, we rejected the flare identification as a spurious result. Intervals discarded in this manner tended to either be made up of a series of small spikes, each marked as an individual flare, but together covering the whole interval, or to be entirely above the global median, thus causing the entire interval to be initially marked as a single flare. Because we did not consider sections where the entire light curve is above the global median to be flares, we are biased against finding flares longer in duration than the observation interval.
Applying this flare identification algorithm yielded 4,672 flares on 1,810 stars. The quiescent flux was then calculated by removing all automatically detected flares and taking the mean of the remaining flux values, with the quiescent flux uncertainty represented by the standard error of the mean. This led to the realization that a large difference between the median and calculated quiescent flux was a good indicator that there was large scale variation in the light curve. 

\begin{figure}[!htb]
  \includegraphics[width=0.63\textwidth]{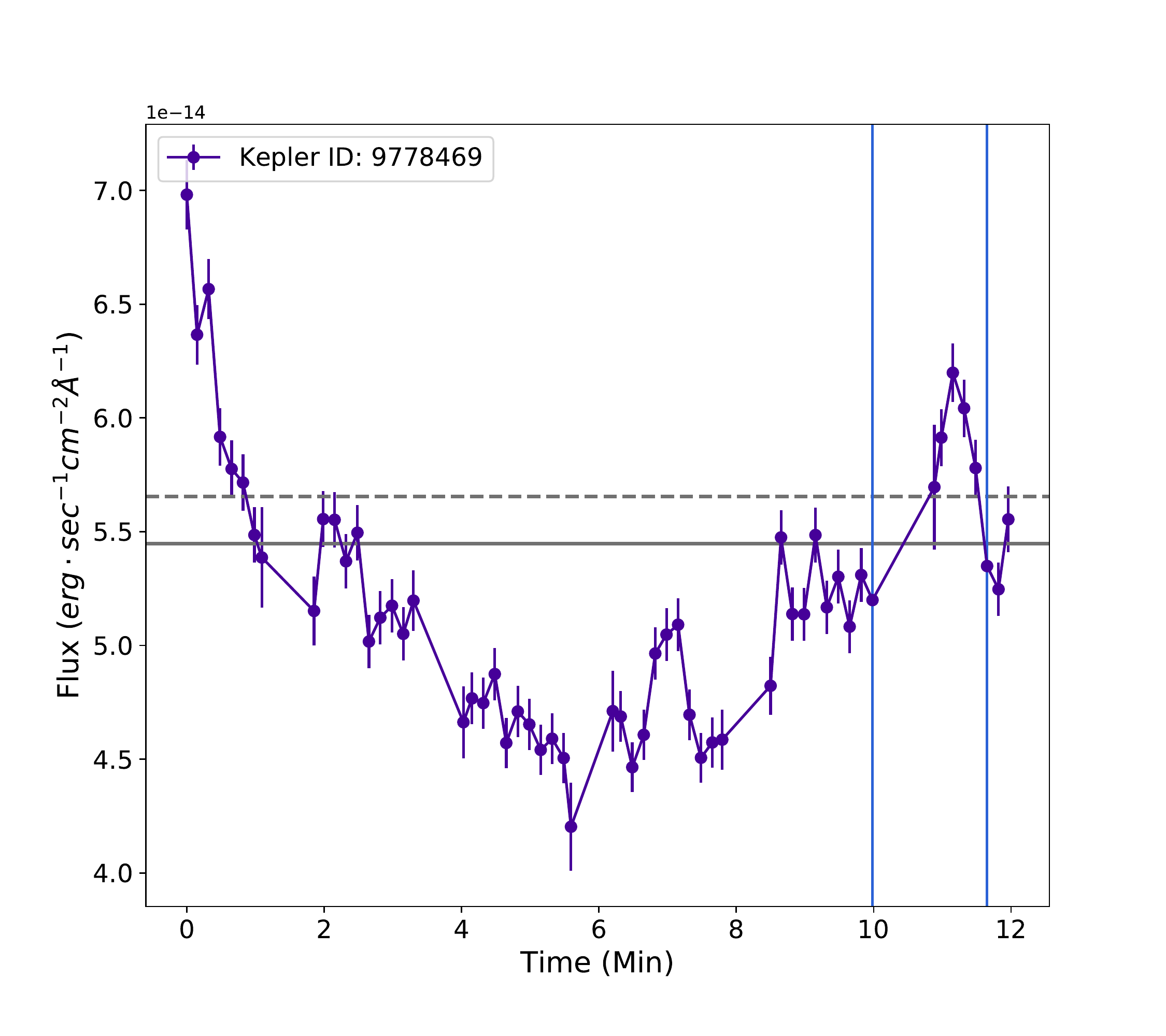}
  \includegraphics[width=0.63\textwidth]{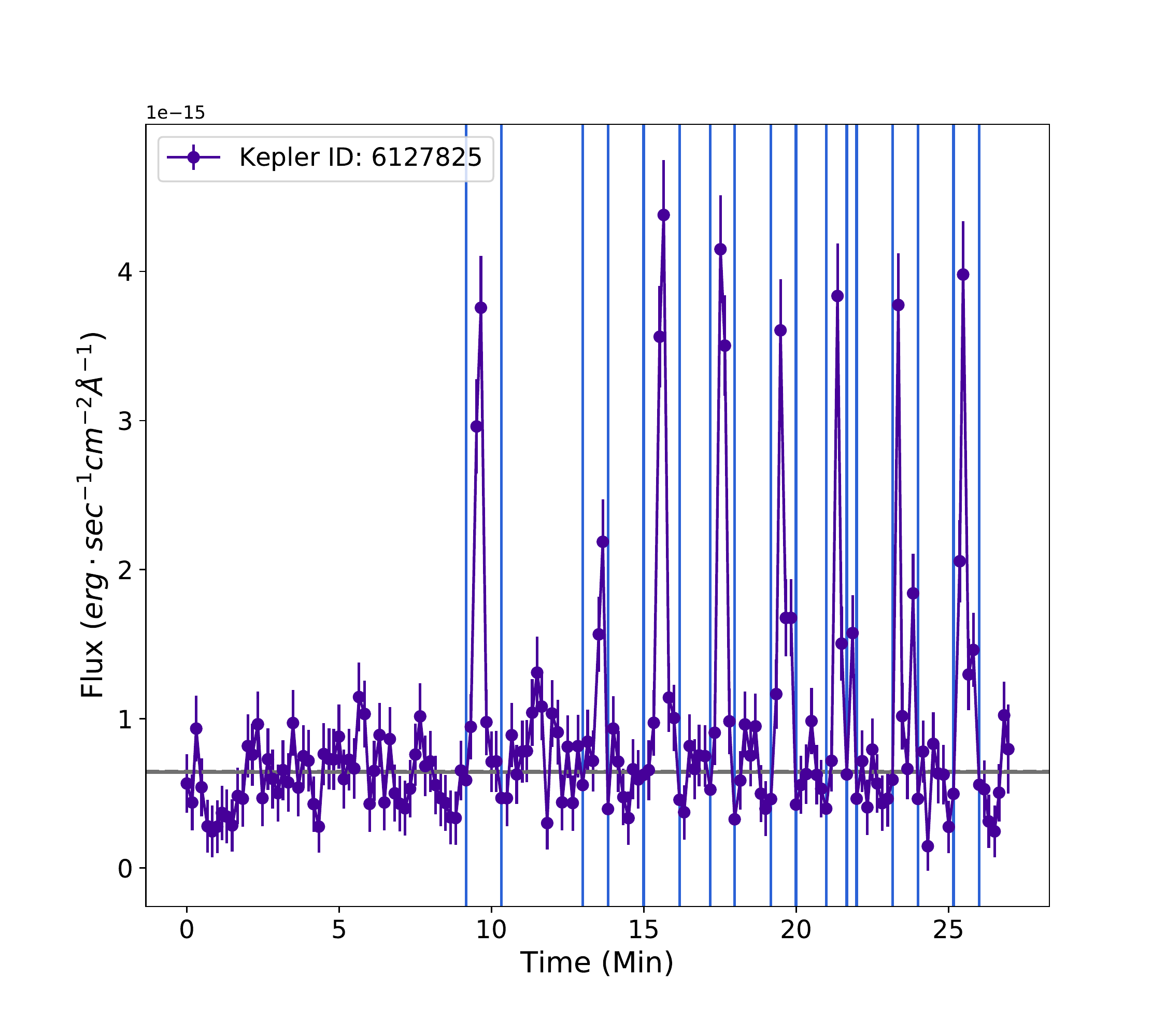}
\caption{Examples of candidate flares which passed screening through Step 4
in flare filtering (Table \ref{tbl:dataRedux}), but which exhibit large- and small-scale periodic variability (top and bottom respectively) and are thus eliminated from the pool of potential flaring stars. Each time bin is 10 seconds, and the error bars are $1\sigma$. The vertical blue lines mark the edges of the putative flares detected by the automated algorithm. The dashed horizontal line marks the global median of the data, while the solid horizontal line is the quiescent flux
In the bottom plot these values are virtually indistinguishable, while in the top plot they are very different.  See text for details.
\label{fig:nonFlares}
}
\end{figure}

Step 5 consisted of manual categorization of all 4,672 automatically detected flares using the Zooniverse platform \citep{Lintott2008}. Putative flares could be marked as false positives, maybes, and true flares,
and were also categorized based on a number of different morphological features.  This allowed us to assess flaws in our automatic flare detection (purposely tuned to be generous to avoid false negatives, at the expense of having to wade through a larger number of false positives) and to keep track of different types of flares. The most common flare misidentifications were due to large- and small-scale periodicity in the light curve.  Large-scale periodicity misidentifications occurred when our algorithm labeled the peaks of large-scale sinusoidal variability as flare activity (see Figure \ref{fig:nonFlares} \textit{top}). These light curves may exhibit flare activity, 
however they need to be detrended before the flare-finding algorithm is run on them. We intend to do this in future, however it has not been done at the time of this writing. Small-scale periodicity misidentifications occurred when  a light curve exhibited a series of small short spikes that were marked as very tiny flares, but occur too regularly and densely to indicate true flares (see Figure \ref{fig:nonFlares} \textit{bottom}).

\begin{figure}[!htb]
  \includegraphics[width=0.49\textwidth]{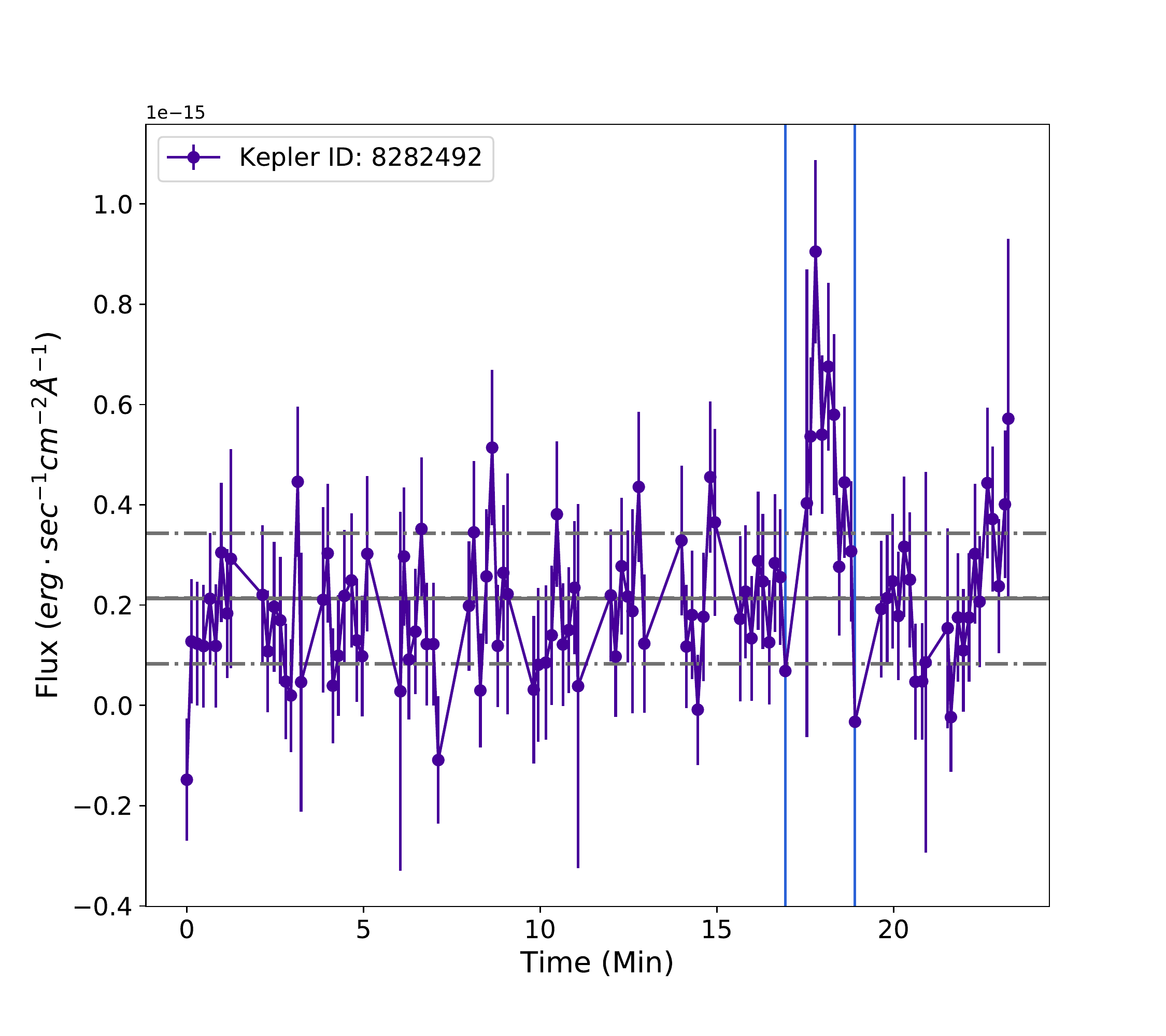}
  
  \includegraphics[width=0.49\textwidth]{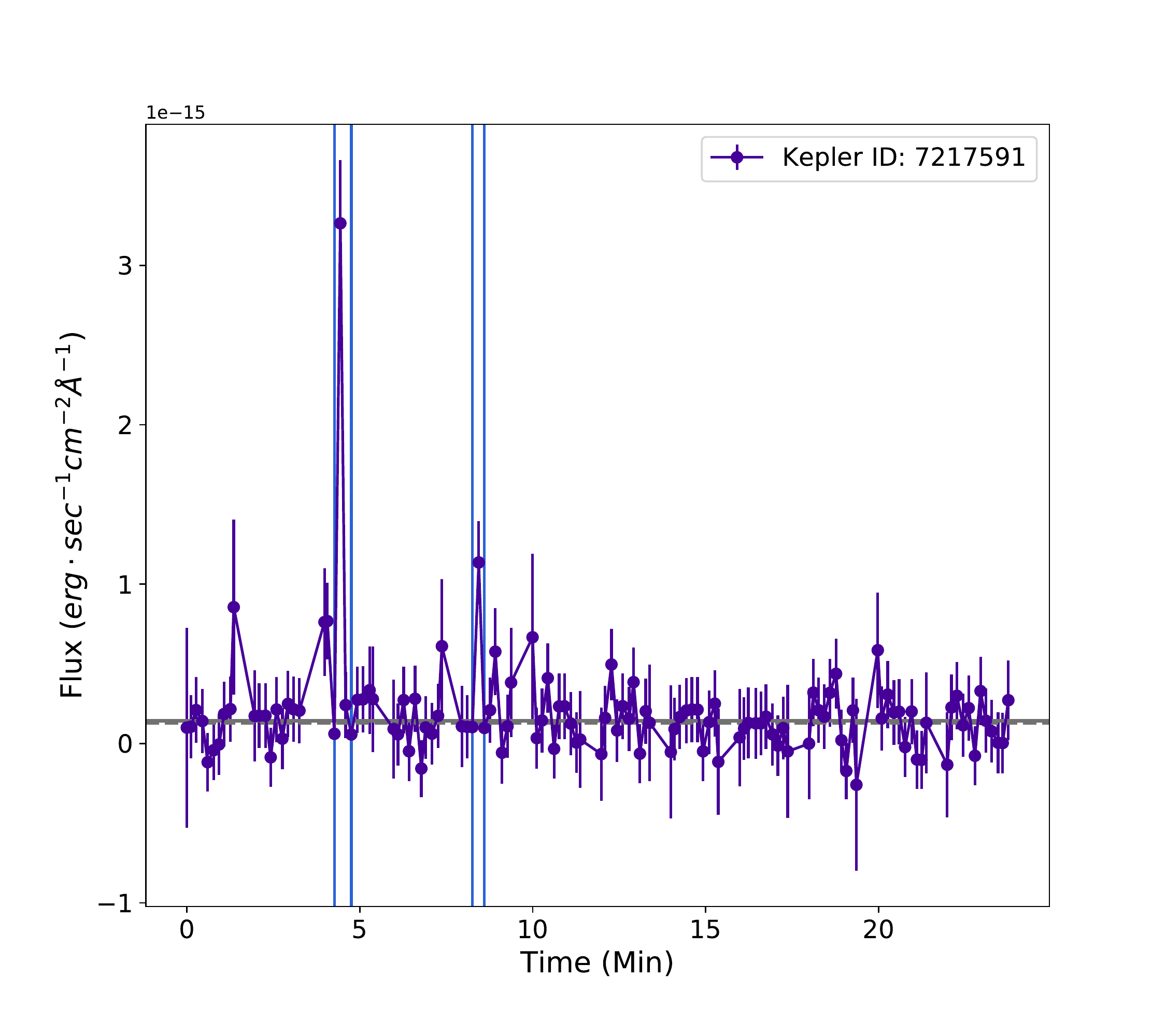}
  \includegraphics[width=0.49\textwidth]{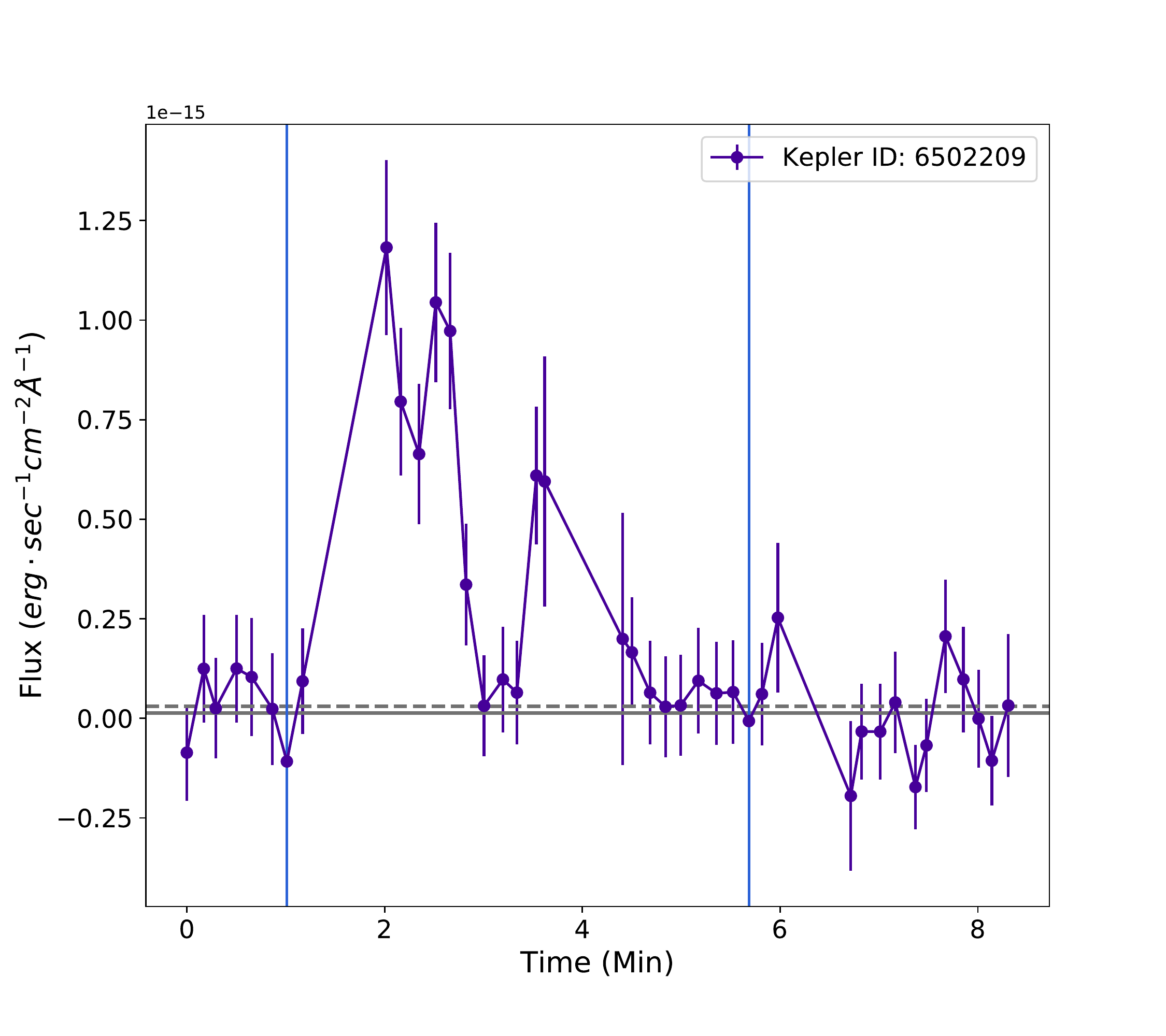}
\caption{Example candidate flares marked ``maybe'' for three reasons. In the top example, the peak enhancements were only slightly larger than the quiescent flux and had a large variation in the quiescent flux level (horizontal dashed lines mark the variance); on the bottom left, candidate flares had very short, often single peaks; and on the bottom right, candidate flares not having the classic fast rise exponential decay shape in combination with large measurement errors. 
Line colors and types are as in Figure~\ref{fig:nonFlares}.
\label{fig:maybeFlares}
}
\end{figure}

A large number of flare candidates were marked ``maybe'' for three main reasons, the most common of which (560 putative flares) was the flare peak being close to the $3.5\sigma$ cut-off when the variance of the quiescent flux was high (see Figure \ref{fig:maybeFlares} \textit{top}). This figure shows the $\pm$1$\sigma$ errors on the quiescent flux measurements. In these light curves, in addition to the putative flares being small, the spread of flux values is high, and there are many measurements that do not rise above the minimum threshold to be considered a flare, but are close. Some of the putative flares marked as such are likely microflares \citep{Robinson1995} while others are statistical noise, and  we did not feel confident distinguishing between the two.  A second reason for a putative flare to be marked ``maybe'' was that it was very short, a single flux measurement elevated significantly above the noise floor (see Fig.~\ref{fig:maybeFlares} \textit{bottom left}). The third main reason was the candidate flare having a shape that deviated significantly from the classic fast rise exponential decay (FRED) shape (see Fig.~\ref{fig:maybeFlares} \textit{bottom right}), usually in conjunction with a large amount of variation in the non-flaring part of the light curve and/or large error bars.
 
\begin{figure}[!htb]
  \includegraphics[width=0.49\textwidth]{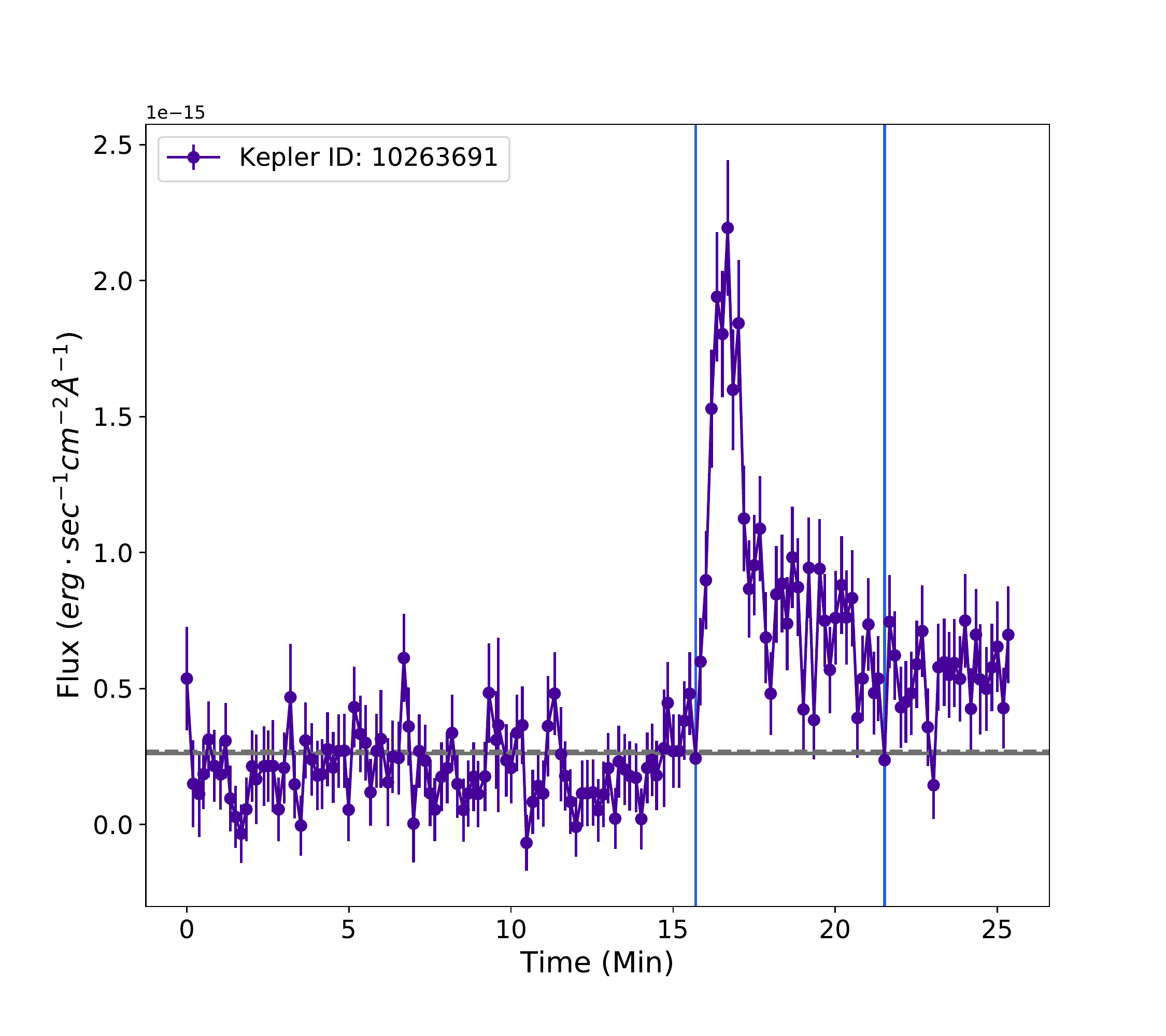}
  
  \includegraphics[width=0.49\textwidth]{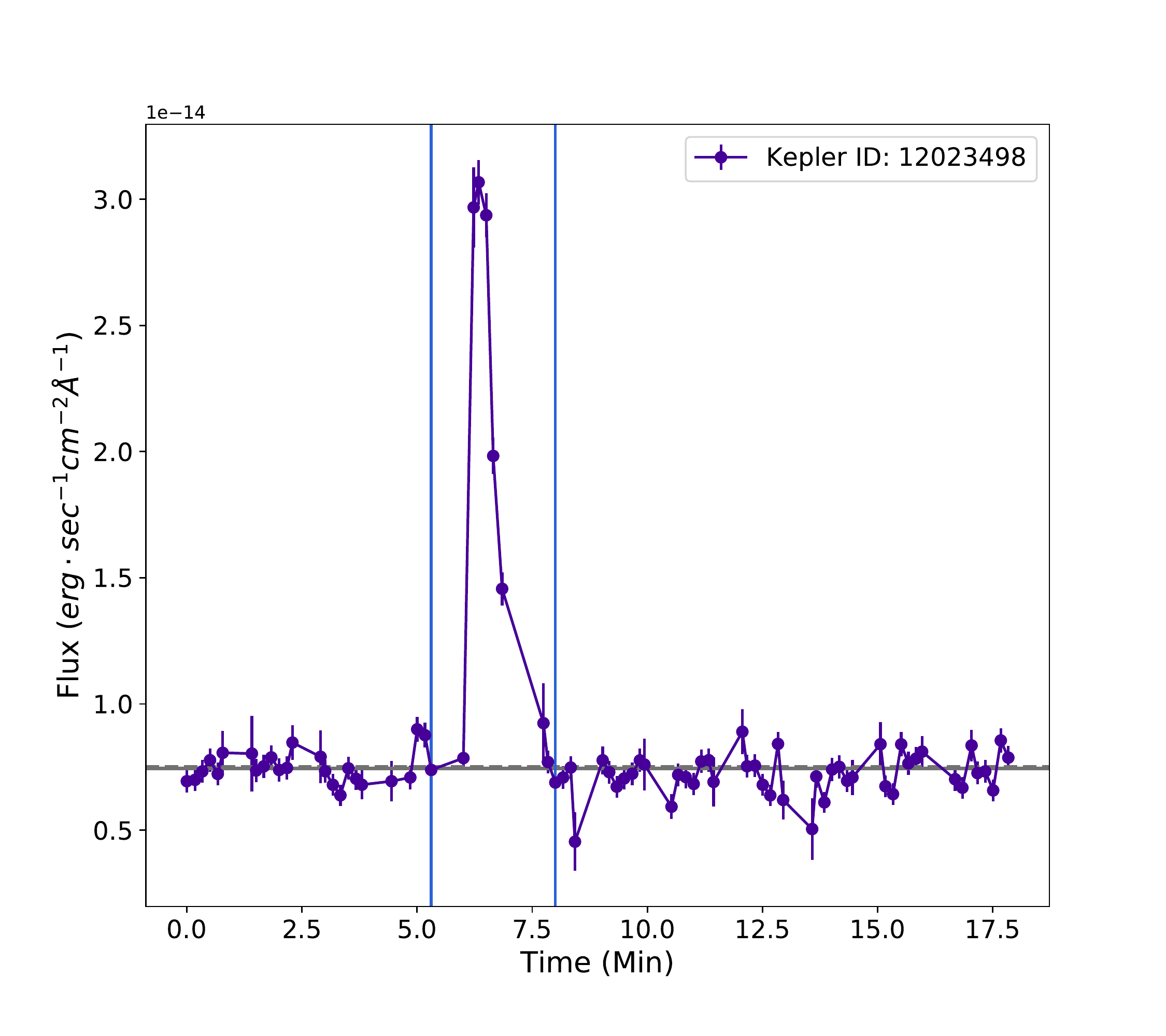}
  \includegraphics[width=0.49\textwidth]{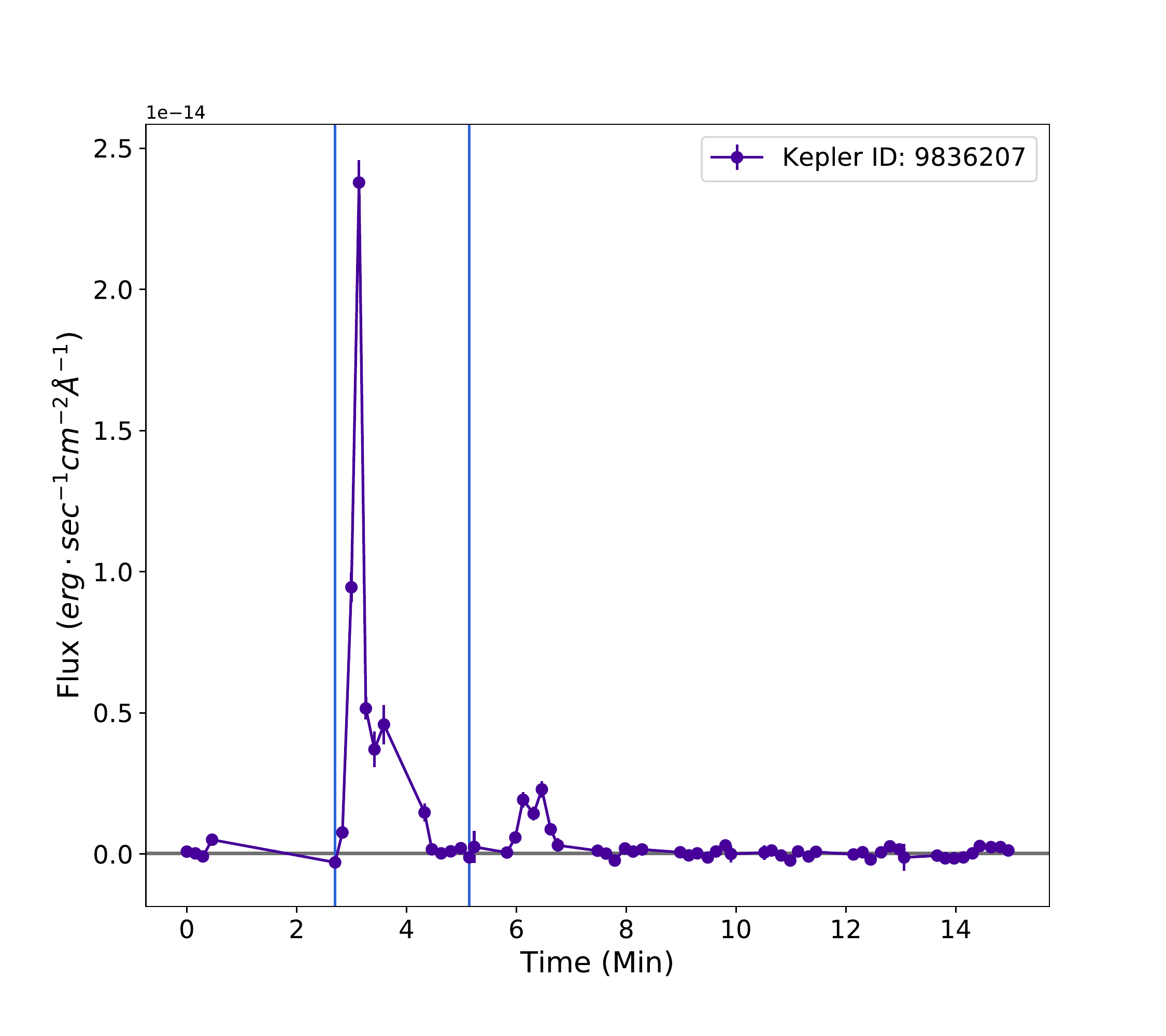}
  \caption{Three examples of GALEX flares exhibiting the classic fast-rise exponential decay (FRED) profile. 
  Line colors and types are as in Figure~\ref{fig:nonFlares}.
  Note that in all three light curves the median and quiescent flux are very similar.
  \label{fig:fredFlares}
}
\end{figure}

\begin{figure}[!htb]
  \includegraphics[width=0.49\textwidth]{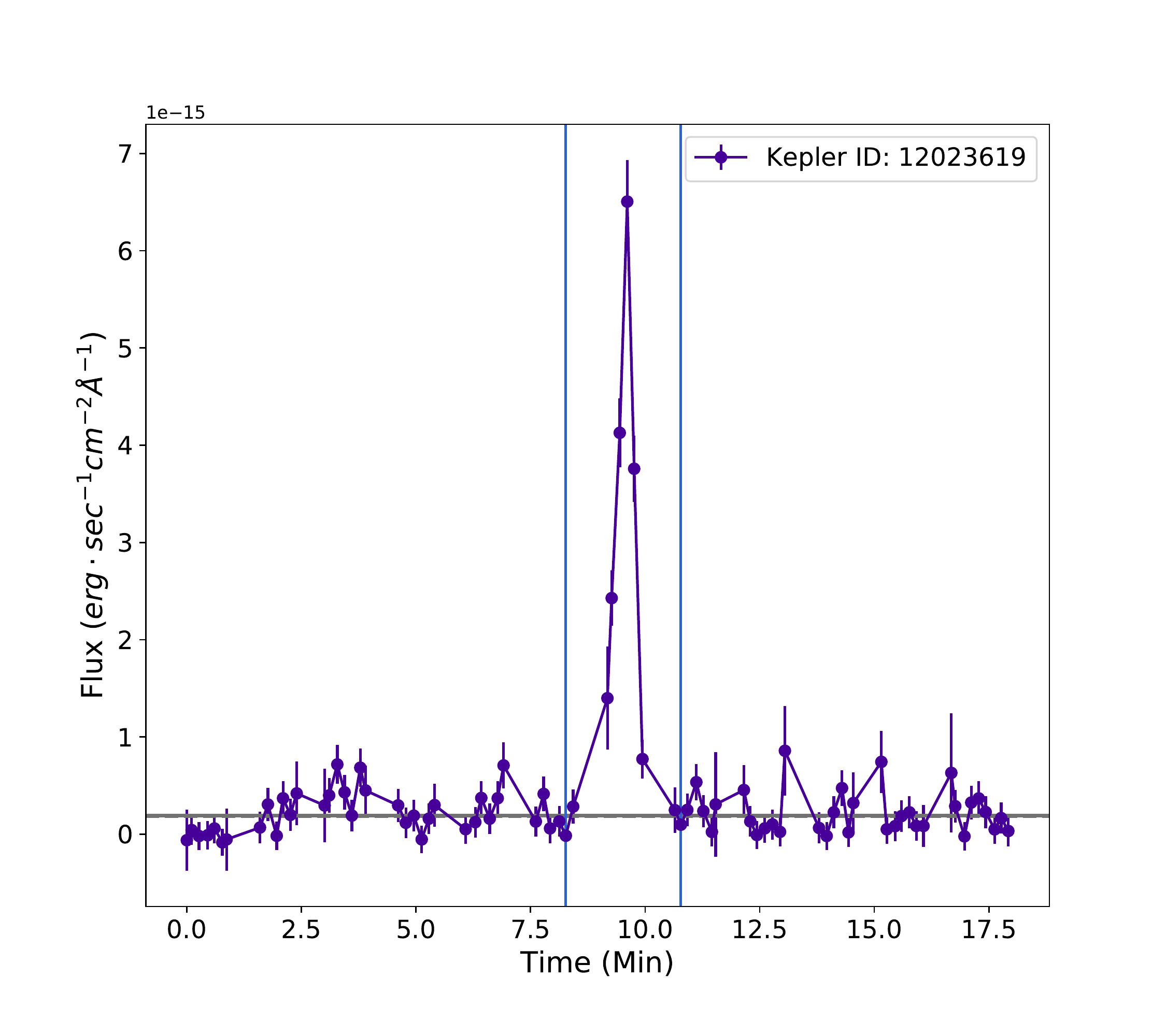}
  \includegraphics[width=0.49\textwidth]{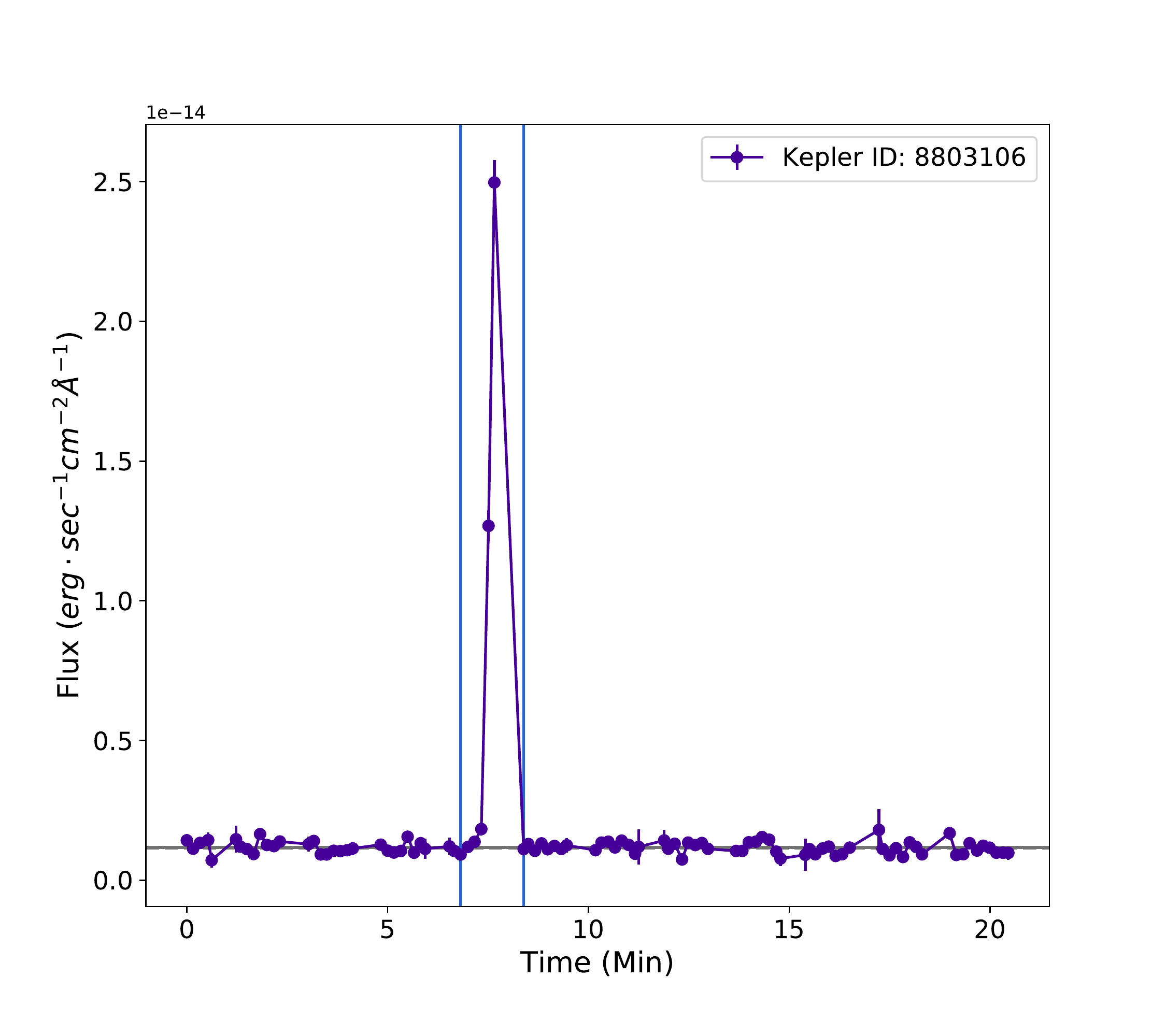}
  
  \includegraphics[width=0.49\textwidth]{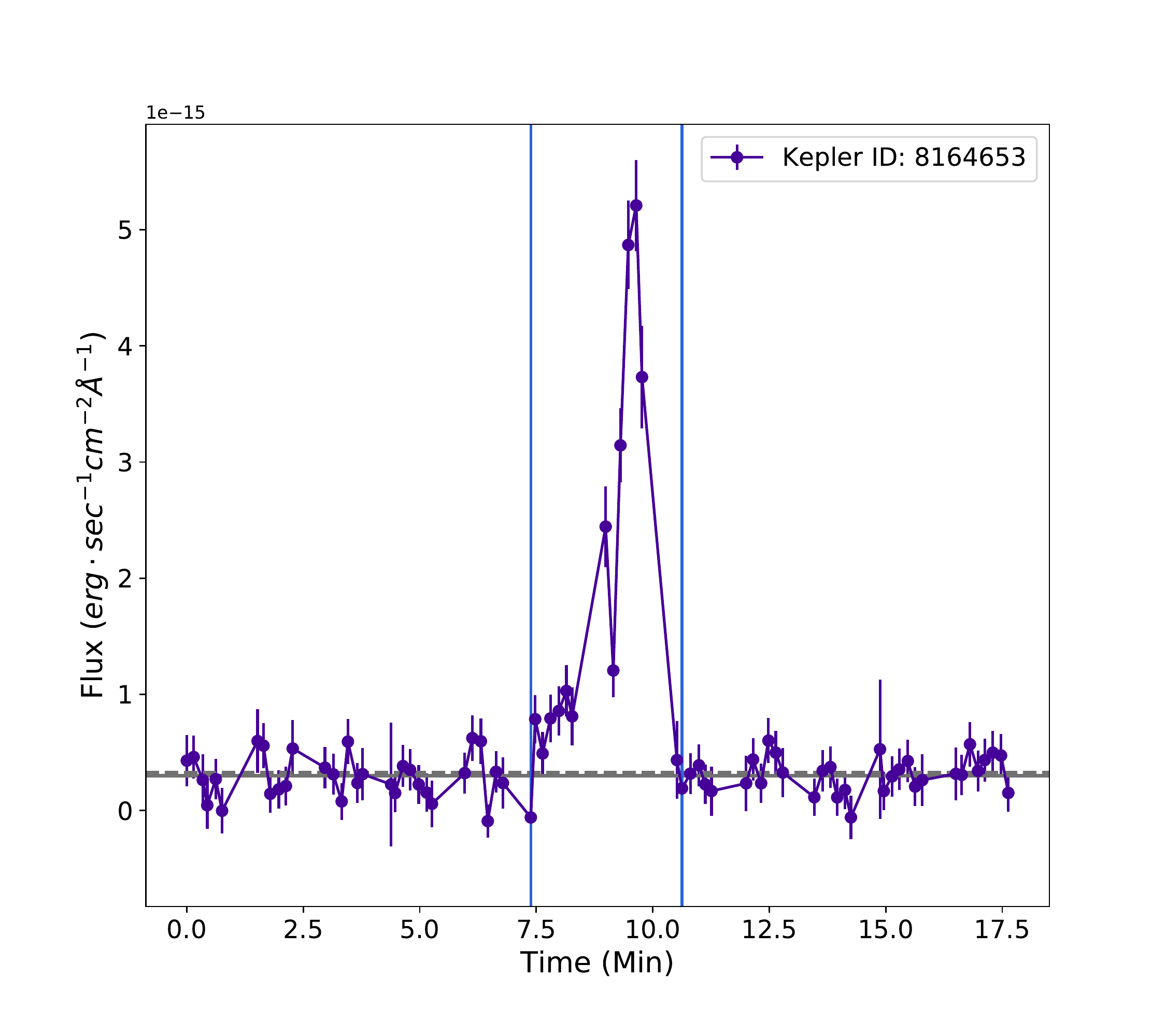}
   \caption{Example GALEX flares exhibiting a range of light curve shapes: the left two light curves appear symmetrical and the third exhibits a reverse-FRED profile. Line colors and types are as in Figure~\ref{fig:nonFlares}.
   \label{fig:symFlares}
}
\end{figure}

\begin{figure}[!htb]
  \includegraphics[width=0.65\textwidth]{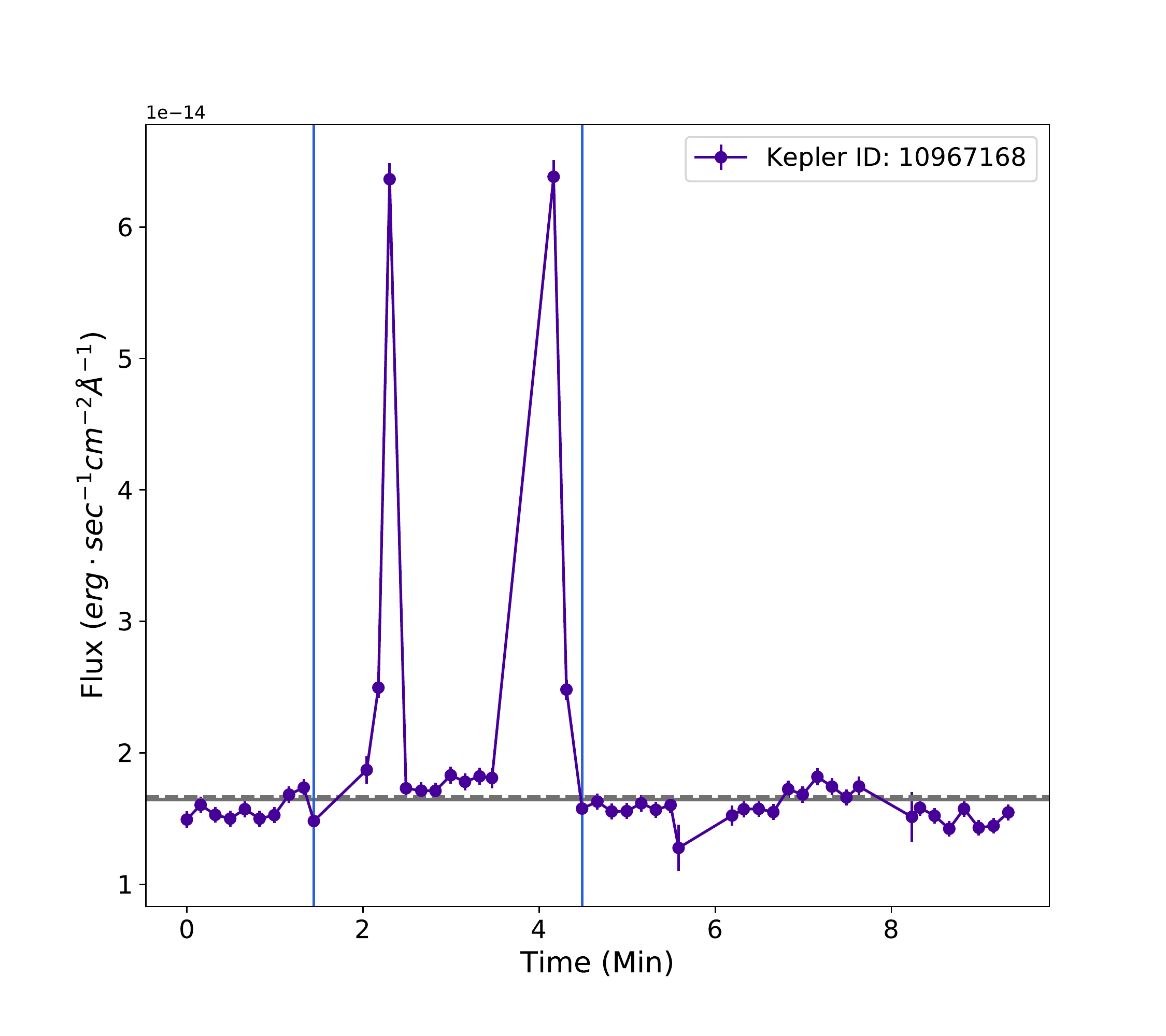}
  \includegraphics[width=0.65\textwidth]{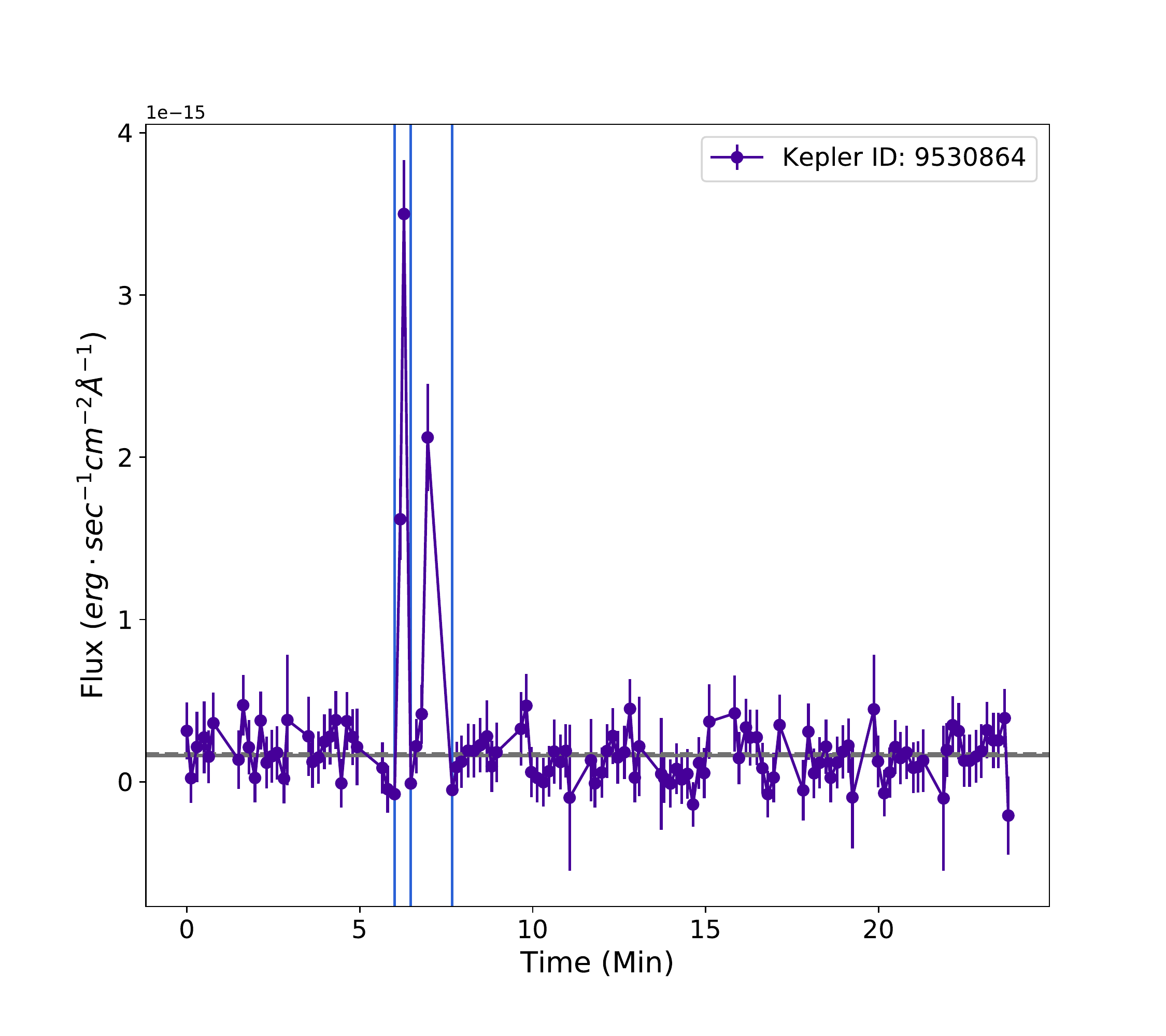}
\caption{Example GALEX flares exhibiting deep subpeaks. In the top example, the algorithm groups what may be two separate flares as one event, while on the bottom, the algorithm identifies two adjacent flares. Line colors and types are as in Figure~\ref{fig:nonFlares}.
\label{fig:multFlares}
}
\end{figure}

In this study we only considered flares manually marked as flaring, thus rejecting the false positives and ``maybe'' flares. This cut our flare count nearly in half, from 4,672 automatically detected flares to 2,194 marked definitely flaring on 1,145 stars through manual inspection. Many of these discovered flares exhibited the classic FRED form (Fig.\ref{fig:fredFlares}), but this was not universally true.  Many flares appeared symmetrical or even seemed to exhibit a reverse FRED-like form (Fig.\ref{fig:symFlares}). Many flares also exhibited sub peaks, some so deep the event was marked as two flares (Fig.\ref{fig:multFlares}).  In total, 860 flares were marked as having sub-peaks, or having nearby neighbors that could be part of the same flare. The bookkeeping for these flares was not altered after inspection and flare confirmation. 

\begin{deluxetable}{lrrr}
  \tablewidth{0pt}
  \tablecolumns{4}
  \tablecaption{Flare Filtering and Data Cuts
    \label{tbl:dataRedux}}
  \tablehead{\colhead{Step \#} & \colhead{Reduction Step} & \colhead{\# Stars} & \colhead{\# Flares}}
  \startdata
  \hline
  \multicolumn{4}{c}{Flare Filtering} \\
  \hline
   1 & Overlapping target lists  & 34,276 & \\
   2 & After first pass flare detection algorithm  &  5,591 & \\
   3 & After second pass flare detection algorithm &  2,356 & \\
   4 & After automatic flare identification &  1,810 &  4,672 \\
   5 & After manual inspection &  1,145 &  2,194 \\
   6 & After color-magnitude cut-off of giant stars &  1,021  &  1,904 \\
   \hline 
   \multicolumn{4}{c}{Data Cuts}\\
   \hline
   7 & Radius and distance determination &  999 &  1,873 \\
   8 & T$_{\rm eff}$ determination & 997 & 1,870\\
   9 & P$_{\rm rot}$ determination & 149 & 276\\
   10 & Cumulative distribution energies (whole flares) & 942 & 1,705\\
  \enddata
\end{deluxetable}


\section{Data Analysis}
The data analysis proceeded in three parts. In the first, we considered the stellar properties of the flaring stars, using this as a final criterion in the flare filtering. Secondly, we determined properties specific to the flares, namely, flare duration, integrated energy, peak flux enhancement. Thirdly, we investigated aggregate flare statistics, using knowledge of both the flare properties and stellar properties.

\subsection{Stellar Population}
\begin{figure}[!htb]
  \includegraphics[width=0.75\textwidth]{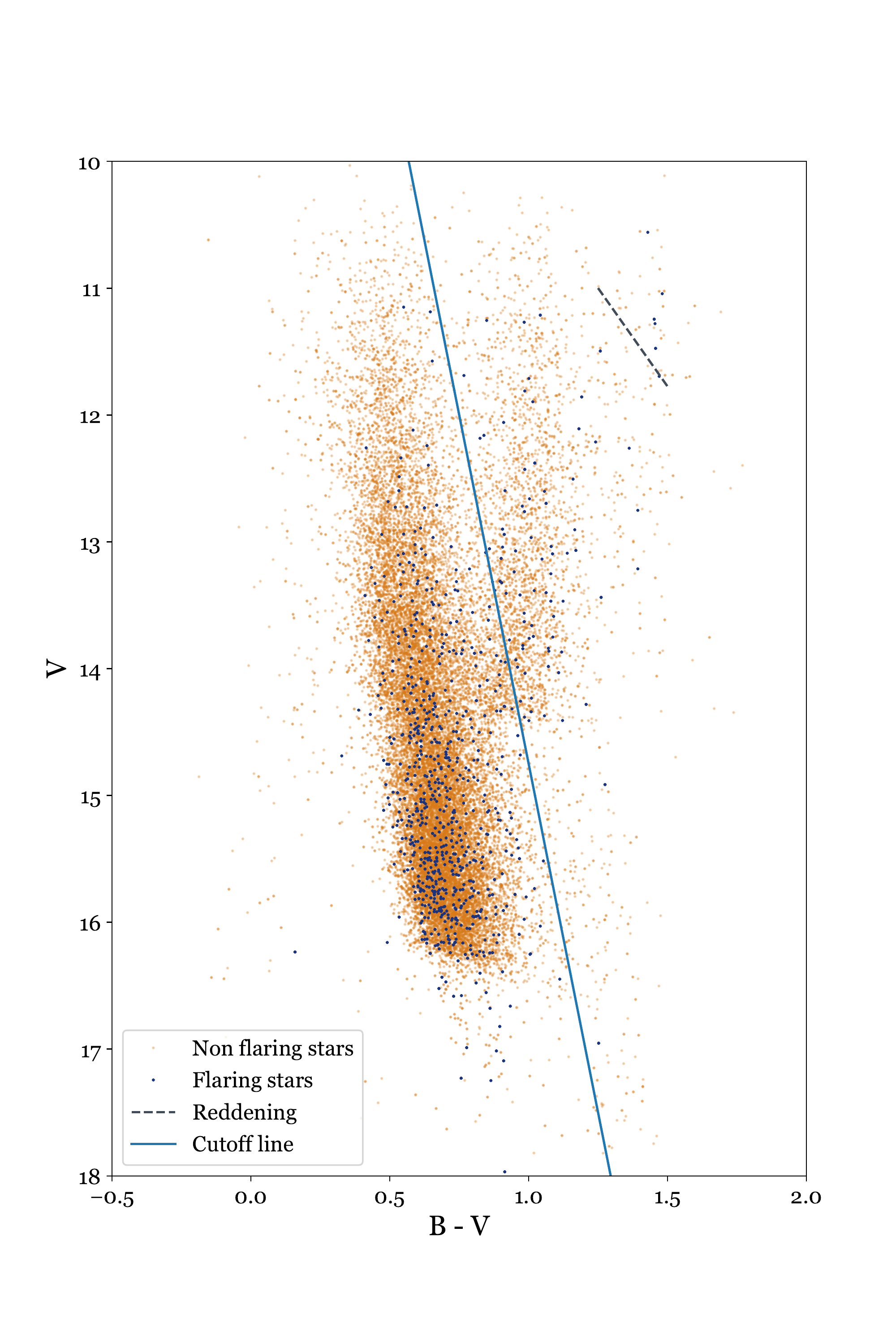}
  \caption{Color-magnitude diagram with flaring stars overlaid on non-flaring stars. Both the B-V color and V-band magnitude originate from the Howell Everett Survey \citep{Everett2012}. The blue line splits the giants from the main sequence stars, and was used as a final cut on the population of flaring stars, so that our sample includes only main sequence stars.  The gray dashed line shows the direction stars would be reddened; we have not applied reddening correction to our points.
    \label{fig:colorMagnitude}
  }
\end{figure}

The target list for the Kepler mission was made from the population of stars brighter than 16th magnitude in the Kepler field of view, optimized for the detection of orbiting Earth-sized planets. The majority of Kepler targets are G-type main sequence stars, with small numbers of M-dwarfs ($\sim3\%$), Giants ($\sim5\%$), and O- and B-type main sequence stars ($\sim0.2\%$) \citep{Batalha2010}. We made use of information in the Kepler Input Catalog \citep{kicref} in our initial reconnaissance of the stellar properties and the final step in the flare filtering process. Specifically, the B-V color and  V-band magnitudes from the Howell Everett Survey \citep{Everett2012} were used to restrict the population studied here to main sequence stars. Figure \ref{fig:colorMagnitude} shows a color-magnitude diagram of both flaring and non-flaring stars after filtering stars up through Step 5 in Table~\ref{tbl:dataRedux}. The line we used as a cut-off between main sequence and giant stars is also shown in figure \ref{fig:colorMagnitude}.  After Step 6 we were left with a  sample of 1,904 flares on 1,021 stars, the vast majority of which are G-type. See Table \ref{tbl:dataRedux} for a complete break down of our flare filtering steps.

\begin{figure}[!htb]
  \includegraphics[width=0.65\textwidth]{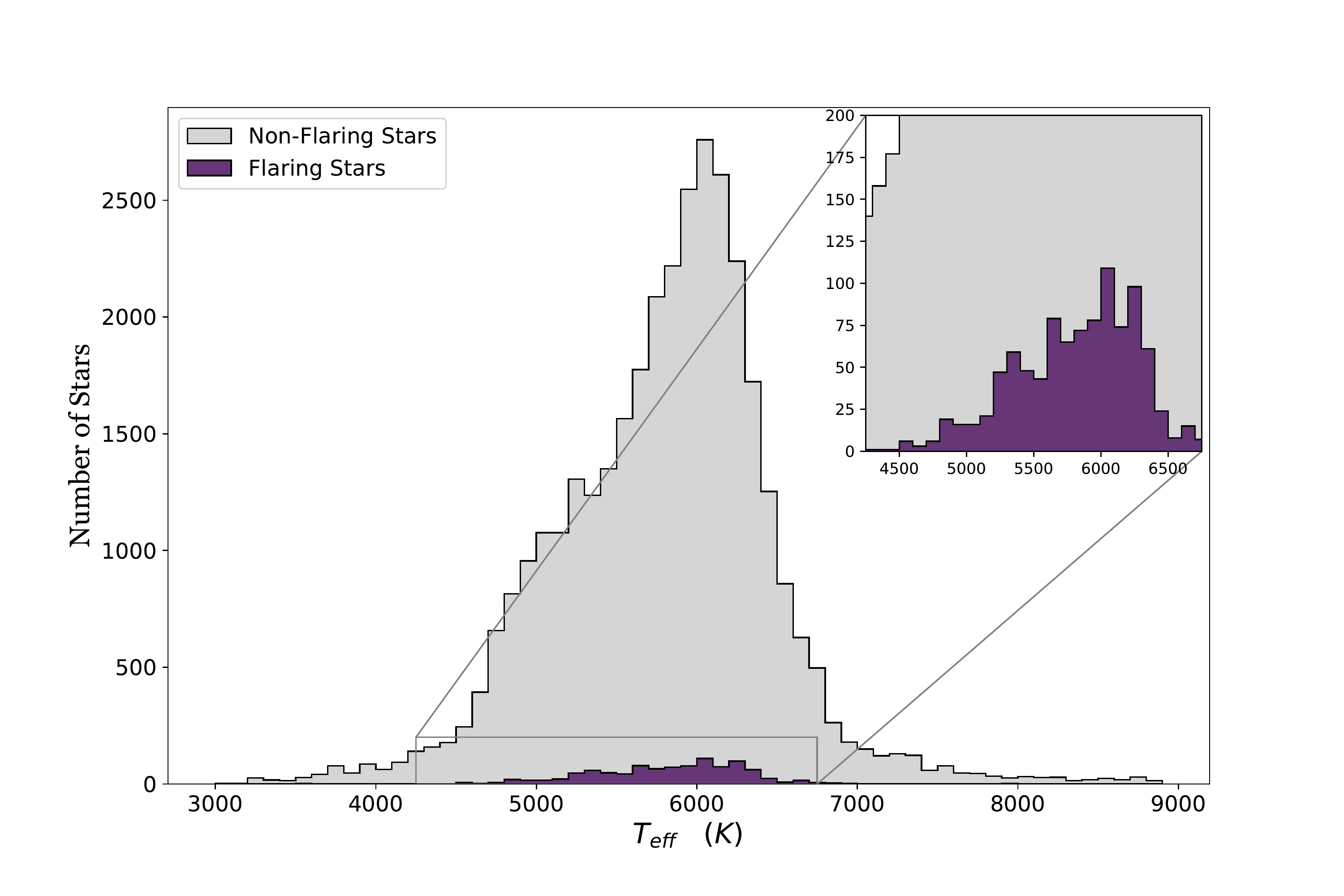}
  \includegraphics[width=0.33\textwidth]{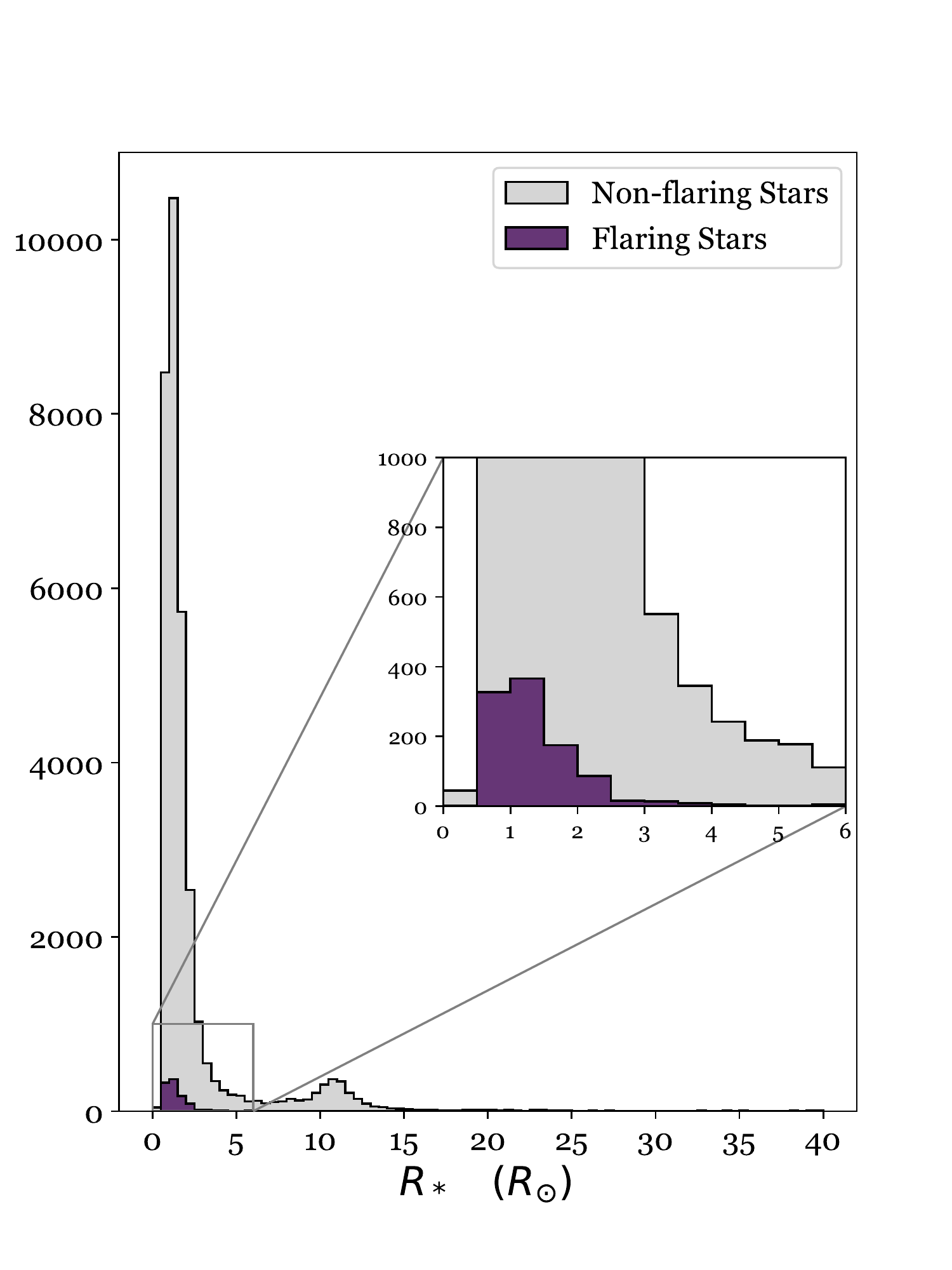}
  \caption{Effective temperature and stellar radius distributions for flaring and non-flaring stars. The effective temperature is the derived effective temperature from the Kepler Input Catalog and is accurate to 200 K. The radii come from \citet{Berger_etAl_2018} and have errors ranging from 2\% to 30\%.
    \label{fig:teff_rad_dist}
  }
\end{figure}

\begin{figure}[!htb]
  \includegraphics[width=\textwidth]{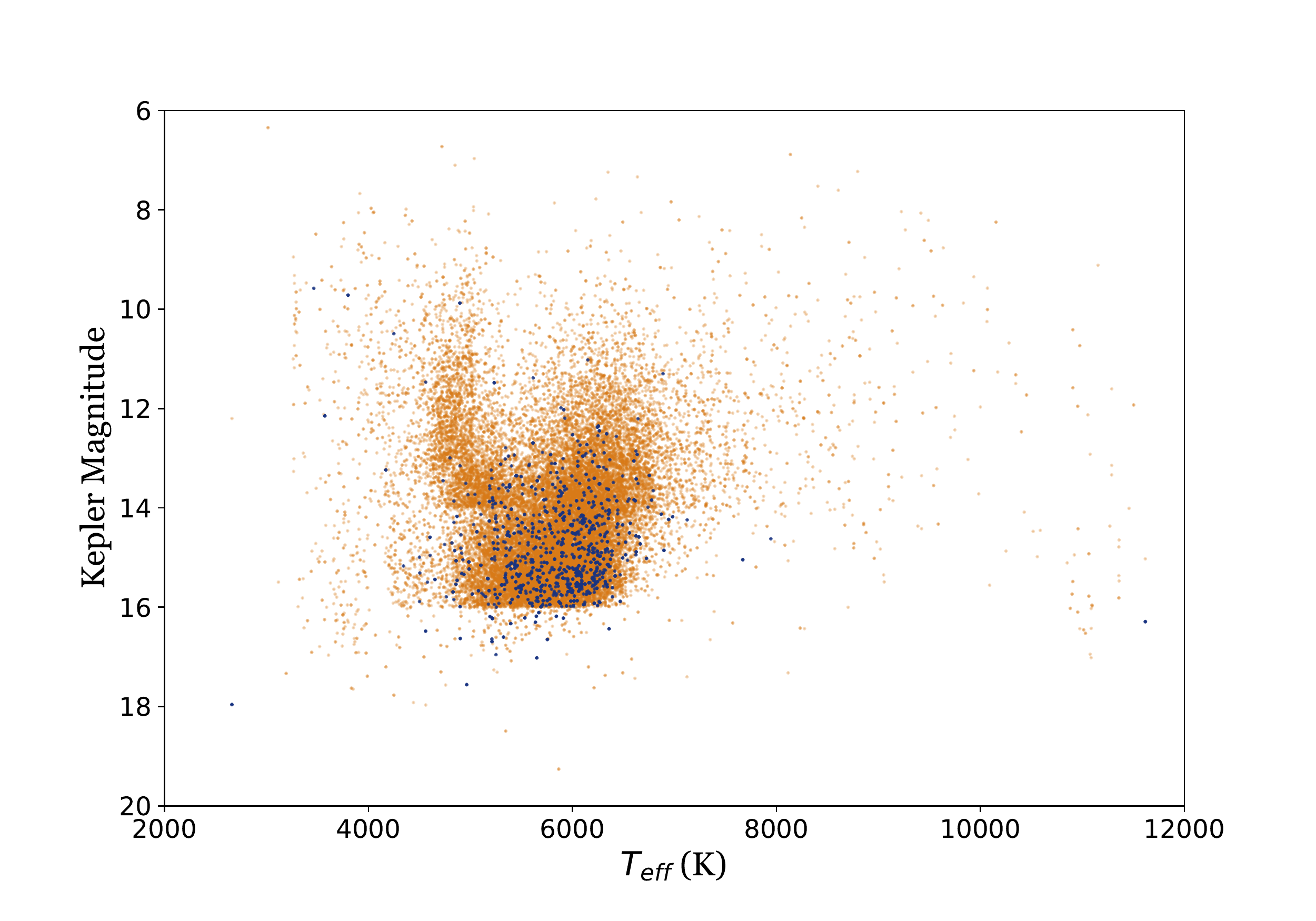}
  \caption{Kepler Magnitude vs. Temperature of flaring and non-flaring stars. Both effective  temperature and Kepler Magnitude estimates come from the Kepler Input Catalog. The effective temperature is the derived effective temperature from the Kepler Input Catalog and is accurate to 200 K. The K\_pmag = 16 cutoff for Kepler targets is clearly evident in this plot; the few stars with fainter magnitudes are due to the Guest Observer program, which did not have a specific magnitude cutoff.
    \label{fig:teff_kepmag}
  }
\end{figure}

As part of our analyses after the flare filtering steps, we utilized additional external sources of information about our flaring stars.  These data cuts are also listed in Table~\ref{tbl:dataRedux}, as we did not have complete overlap between these and our sample.  Figure \ref{fig:teff_rad_dist} shows the effective temperature ($T_{\rm eff}$) and radius ($R_*$) distributions for flaring vs non-flaring stars after the final step in flare detection, described above. The effective temperature estimates come from the Gaia DR2 catalog, and have errors from 1-15\%. The majority of stars, 968 out of the 1,021, had T$_{\rm eff}$ determinations. We use \citet{Berger_etAl_2018}'s stellar radius values using the Gaia DR 2 data release\footnote{http://doi.org/10.17909/t9-bspb-b780}, for which the overlap was also very large, 999 stars out of our 1,021 flaring stars. The majority of the flaring stars are around the same size as our sun, running from smaller than the sun to about $6 R_{\odot}$.  The largest flaring star is $\sim15 R_{\odot}$, and the smallest is $\sim0.5 R_{\odot}$. The bulk of the flaring stars however have radii between $0.7  R_{\odot}$ and $6  R_{\odot}$. The errors on radius range from 2\% to 30\%, and the errors in distance range from 0.3\% to 40\%. The plot of Kepler magnitude (K$_{p}$) vs. $T_{\rm eff}$ (Fig. \ref{fig:teff_kepmag}) shows more clearly how the flaring stars cluster in the 5000-6000 K range.  Also visible in this plot is  a sharp cutoff below K$_{p}$=16 due to the Kepler selection criteria \citep{Batalha2010}. The existence of fainter targets comes from Guest Observer pointings. 

\begin{figure}[!htb]
  \includegraphics[width=0.8\textwidth]{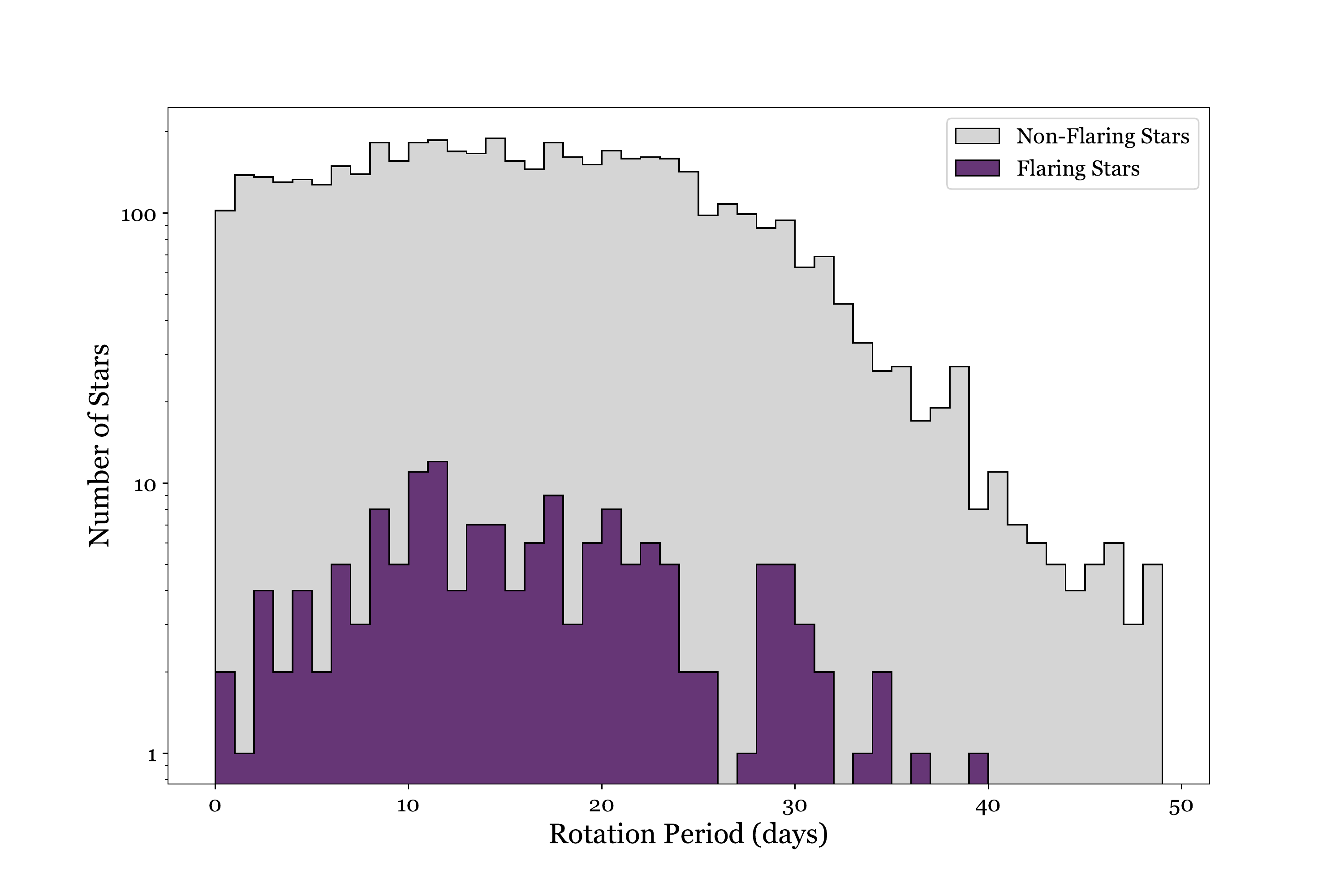}
\caption{Histogram showing the rotation periods of flaring versus non-flaring stars
taken from \citet{McQuillanEtAl2014}.  The rotation periods of the flaring and non-flaring stars follow similar trends, with the flaring stars having a shorter upper limit period than the non-flaring stars.  Only a small number of stars in our sample
have rotation period measurements; see text for details.
\label{fig:rotationHist}
}
\end{figure}

\cite{McQuillanEtAl2014} determined rotation periods of main-sequence stars in the Kepler field using the Kepler light curves. A total of 66\% of our stars (flaring and non-flaring) appear in this catalog, however only 14\% of the stars have defined periods (using the ``Periodic Table'' of rotation periods). Only about 200 of the flaring stars have measured rotation periods, with the periods ranging from a third of a day to 66 days. Figure \ref{fig:rotationHist} displays a histogram of the rotation periods of both flaring and non-flaring stars.

\subsection{ Flare Properties}
The main flare properties to be determined from the light curves are the flare duration, the integrated energy, and the flare peak flux enhancement. 

\subsubsection{Peak Flux Enhancements}
The flare peak flux enhancement is one of the outputs of the flare-finding algorithms described in \S 2, and is quantified either as the number of sigma above the median, or in the ratio of peak flux to quiescent flux. The right panel of Figure~\ref{fig:flareDurPeakFlux} displays the distribution of peak flux enhancements. The majority of the flares (60\%) have peak flux enhancements between $3.5\sigma$ (the cutoff for being considered a flare) and $6\sigma$, with the largest being $48\sigma$. The values of flare flux enhancement to quiescent flux ratio range from 1.5 to 1700. For a small number of stars, the quiescent flux is very close to the background flux, so the peak/quiescent enhancement may actually be higher than calculated. 

\subsubsection{Flare Durations}
\begin{figure}[!htb]
  \includegraphics[width=\textwidth]{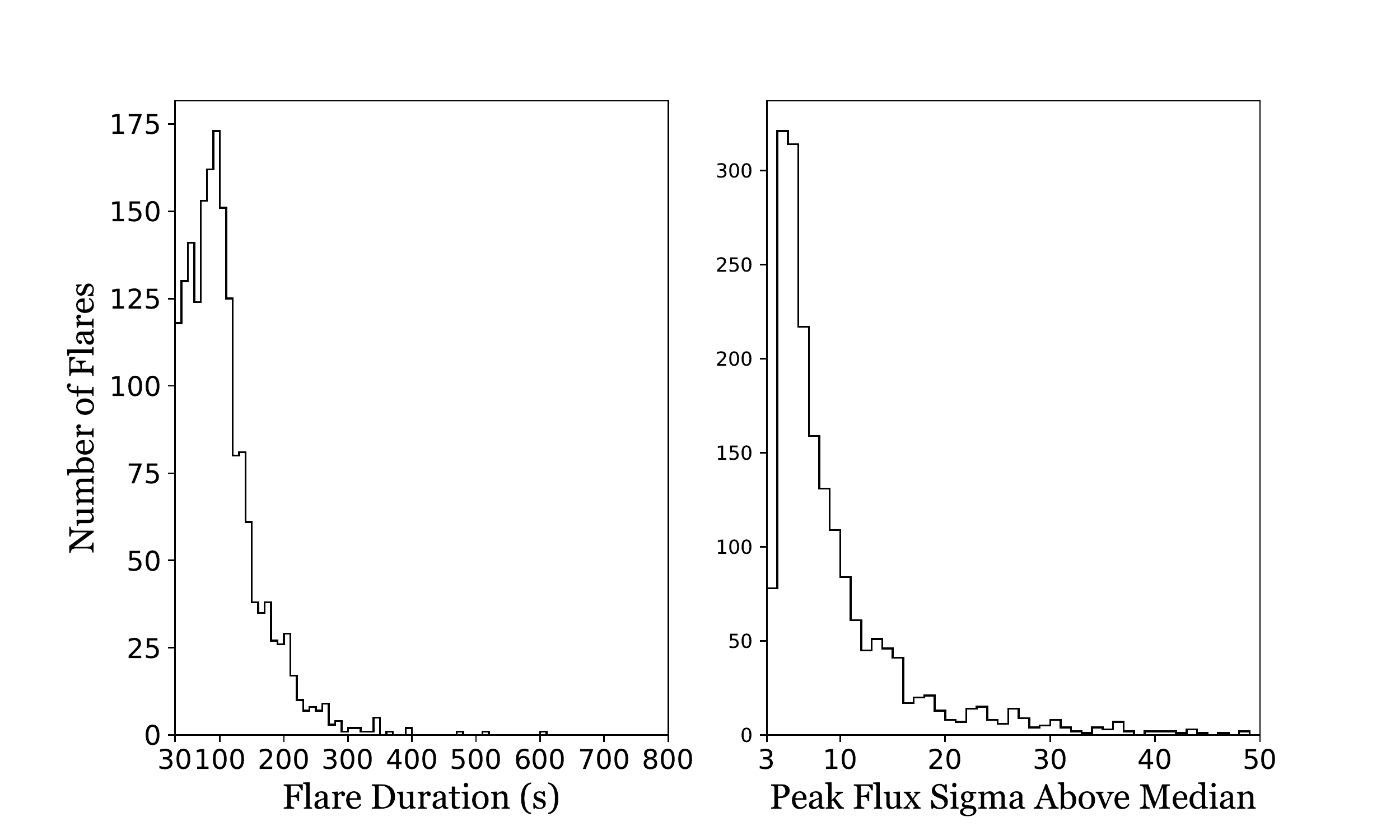}
\caption{Histograms showing the distribution of flare duration (left) and peak flux (right). The bulk of the identified flares are only a couple of minutes in duration, with nearly all of them (94.5\%) under 5 minutes.  Most of the flares are also on the small side in terms of maximum flux, with majority (60\%) being between the minimum peak flux of  $3.5\sigma$ and $6\sigma$.  
\label{fig:flareDurPeakFlux}
}
\end{figure}

Durations are determined from the flare-finding algorithms, also described in \S 2; they are the stop minus start times of the flare, taken from the first points on either side of the peak less than or equal to the global median, or the edge of the interval, whichever was reached first. Flares which extend beyond the edge of an observation interval are "cut off flares" and illustrated in plots with arrows indicating that the calculated duration is a minimum duration. These were excluded from calculation of flare energies. The bulk of identified flares are very short in duration, with 90\% of them under 4 minutes (left panel of Figure \ref{fig:flareDurPeakFlux}), 94.5\% under 5 minutes, and 98\% under 9 minutes. 
 
\begin{figure}
    \centering
    \includegraphics[scale=0.5]{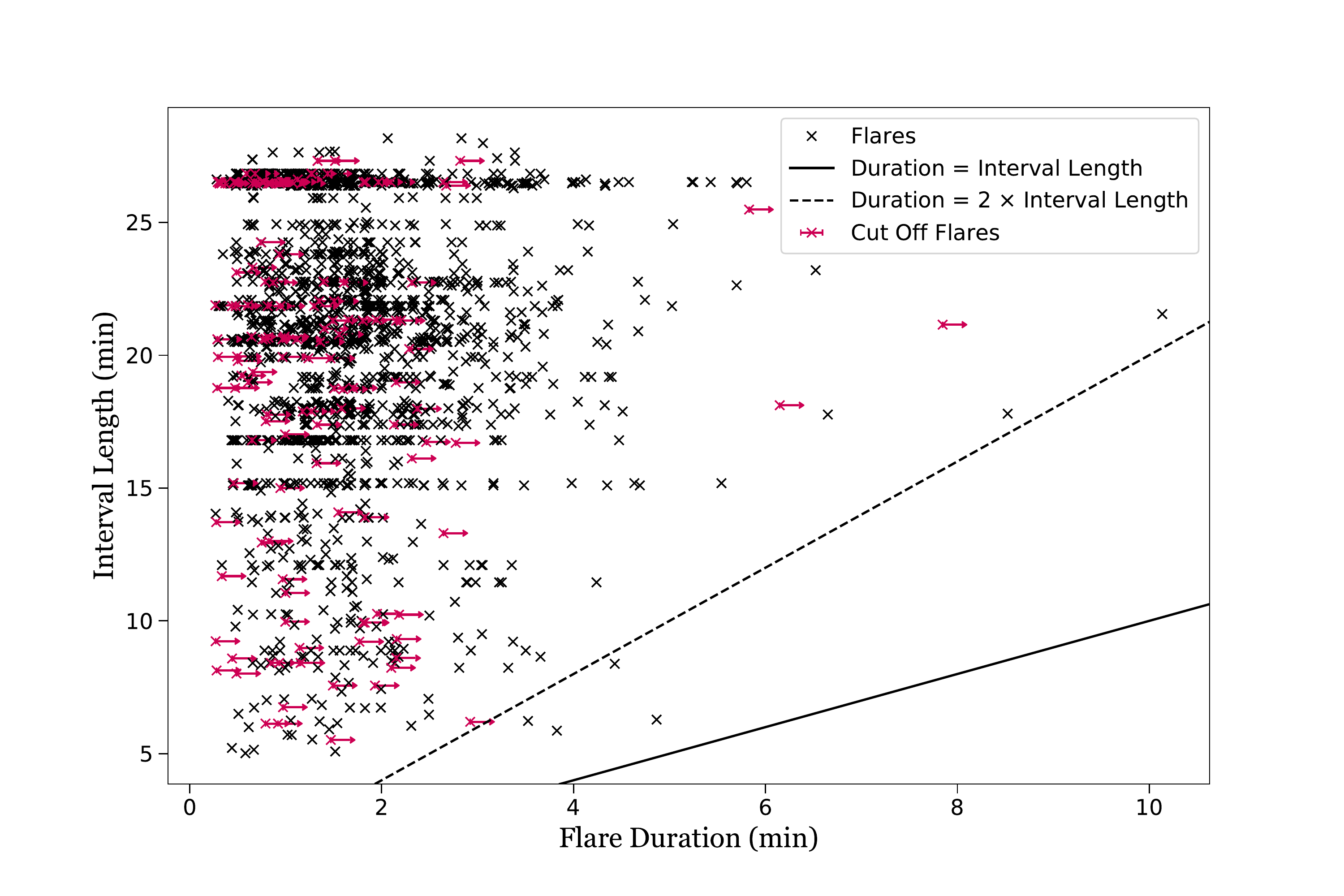}
    \caption{Scatter plot of flare duration and corresponding interval length. Flares that extend past the observation interval are marked with red arrows. The solid line represents equal flare duration and interval length, while the dashed line shows interval lengths of twice the flare duration. The bulk of the flare durations are less than five minutes, and are not affected by the observation interval. }
    \label{fig:flareDurIntLength}
\end{figure}

The characteristics of the GALEX data collection, being based on individual time intervals of 5-30 minutes, biases the study to finding short flares.  The scatter plot of interval length in which the flare was located vs. the flare duration, Figure~\ref{fig:flareDurIntLength}, reveals no effect of interval length on the flare duration which means that we are seeing a good representation of the flares present within our sensitivity range ($\sim$ 0.5-20 min).
 
\begin{figure}[!htb]
  \includegraphics[scale=0.5]{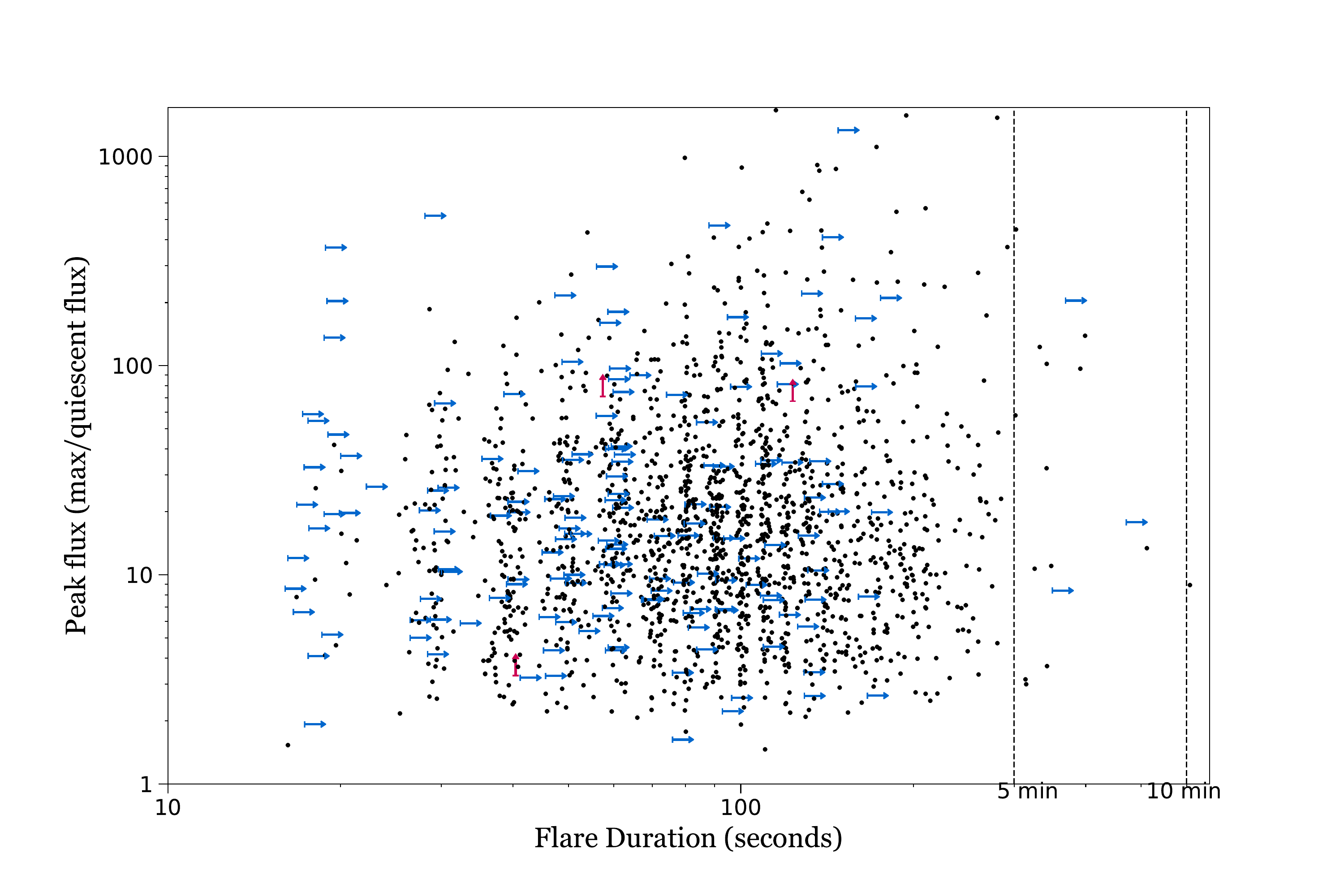}
\caption{A scatter plot of peak flux vs.flare duration, arrows indicate when the calculated value is a minimum, blue horizontal for flares that extend past the observation interval, and red vertical for stars with measured quiescent flux comparable the background flux. There does not appear to be a strong correlation between flare duration and maximum enhancement. The vertical clustering is due to the 10 second binning of the light curves, which artificially bunches up the flare durations in 10 second intervals. The solid vertical line at 5 minutes indicates the minimum required continuous interval length in our data selection; see \S 2 for more details. 
\label{fig:durationVsMaxFlux}
}
\end{figure}

Figure \ref{fig:durationVsMaxFlux} plots peak relative flux (max/quiescent flux) against flare duration (as calculated by our flare-identification algorithm). The cut off flares are shown as blue arrows indicating that the calculated duration is a minimum duration, and the flares with a lower limit on peak/quiescent enhancement have vertical lower limits indicated. Again it is clear that our flare population is small and short.  Additionally there is not a marked trend towards higher enhancements on longer flares, although the range of peak fluxes increases with longer durations. In Figure \ref{fig:durationVsMaxFlux} there are visible vertical lines along which flare durations cluster; this is an artifact of our 10 second binning which causes flare durations to artificially be distributed in 10 second intervals.  Furthermore, true flare duration is not extrapolated for cut off flares (these flares are marked with blue arrows), so there is an artificial upper limit on detectable flare length ($\sim30$ min, or 1800 sec) in this dataset, even if detected flares do extend beyond the observation limits. As mentioned in \S 2 our sample of flares include flares with deep subpeaks that might be more accurately marked as separate flares, as well as sets of neighboring flares that might more accurately be considered single flares with subpeaks. While we did not update the bookkeeping of these flares, with 326 flares showing subpeaks, and 492 having nearby neighbors, we believe that the mis-categorisations in either direction likely balance out and do not affect our overall statistics.

\subsubsection{Flare Energies}
A fluence $F$ is first calculated by subtracting the quiescent flux from each point in the flare light curve, summing over all time bins in the flare, and multiplying by the bandpass of the instrument, \\
\begin{equation}
F = \sum\limits^t (f_t-f_{qui})\omega_{galex}\Delta t \;\; {\rm erg \; cm^{-2}}.
\end{equation} 
Here $f_{qui}$ is the quiescent flux, $f_t$ is the flare flux in time bin t, $\Delta t$ is the width of time bin t, and $\omega_{galex}$ is the equivalent width (FWHM) of the GALEX waveband in \AA. We used the value $\omega_{galex}$ = 795.65\AA\ \footnote{http://svo2.cab.inta-csic.es/svo/theory/fps3/index.php?id=GALEX/GALEX.NUV} \citep{Rodrigo_2012ivoa.rept.1015R}. The total radiated energy in the GALEX NUV
bandpass is then determined via\\
\begin{equation}
 E_{\rm NUV} = 4 \pi d^2 F \;\; {\rm erg},
\end{equation}
where $d$ is distance to the star. Fluence and energy calculations are only
performed for flares with a calculated duration and enhancement over quiescent emission. We used stellar distances calculated by \citet{BailerJones2018} using the Gaia DR2 data.  Error bars on the integrated energy are propagated from the 68\% confidence interval high and low bounds of the distances, one standard deviation errors on the flux, and standard error of the mean on the quiescent flux calculation. Additionally we compensated for reddening using the mass extinction coefficient $R(NUV)=7.06\pm0.22$ given in \citet{Yuan_etal_2013} Table 2, and individual reddening values calculated with the dustmaps package \citep{green_2018}.

The flare energies were calculated initially for the GALEX NUV bandpass. We calculated the energy partition using the calculations in \citet{ostenwolk2009} to account for the fact that a flare will produce radiation across a range of wavelengths, not just those under consideration here. We calculated the flare energy partition as follows:\\
\begin{equation}
    p_{bol} = \frac{E_{\rm NUV}}{E_{\rm bol}} = \frac{E_{\rm NUV}}{E_{\rm cont}}\left(\frac{E_{\rm cont}}{E_{\rm bol}}\right) \;\;,
\end{equation} 
where $p_{\rm bol}$ gives the fraction of bolometric flare energy that shows up in the NUV bandpass; $E_{NUV}$ is the radiated flare energy in the GALEX NUV bandpass, $E_{bol}$ is the bolometric flare energy (the flare energy radiated over all wavelengths), and $E_{\rm cont}$ is the radiated flare energy in the hot continuum. In using this calculation we assume that the dominant energy contribution in the GALEX NUV bandpass is hot blackbody radiation, which is a fraction of the bolometric radiated energy. The GALEX NUV bandpass is $1771-2831$\AA  \citep{Morrissey_2007}. We use  the hot blackbody ($1400-10000$\AA) from \citet{ostenwolk2015} Table 2 as the wavelength range over which to estimate the hot blackbody contribution. We note that recent results of \citet{kowalski2018} have pointed out that the blackbody assumption may underestimate the energy release in the NUV by factors of 2-3. Because energy is proportional to intensity, we  use integrated intensity to calculate the energy ratios: \\
\begin{equation}
\frac{E_{\rm NUV}}{E_{\rm cont}} = 
\frac{\int_{\rm NUV_{1}}^{\rm NUV_{2}} I_{BB}(T,\lambda)d\lambda}{\int_{\rm cont_{1}}^{\rm cont_{2}} I_{BB}(T,\lambda)d\lambda} = 0.219 \;\;,
\end{equation}
where $I_{\rm BB}(T,\lambda)$ is the Planck function
at temperature $T$ and wavelength $\lambda$, the temperature of the blackbody used for the hot blackbody calculation is 10,000 K, the empirically determined temperature of the plasma emitting a white light flare \citep{Kowalski2013}. We use $E_{\rm cont}/E_{\rm bol} = 0.6$ from \citet{ostenwolk2015}'s Table 2, which results in $p_{bol}  = 0.132$. All the energies shown in plots in this paper have been corrected to a bolometric energy value; multiplication by this value of $p_{\rm bol}$ will give the radiated energy in the GALEX NUV bandpass. A selection of our flares and associated properties are shown in Table \ref{tbl:catalog}. 
  
\begin{deluxetable}{cccccccccccccccc}
\rotate
\tabletypesize{\tiny}
  \tablewidth{0pt}
  \tablecolumns{11}
  \tablecaption{All detected and vetted flares with calculated energies. This table includes 1,790 flares that do not extend past the observation interval and are on stars whose quiescent flux is distinguishable from the background flux. This table is published in its entirety in the machine-readable format. A portion is shown here for guidance regarding its form and content.
    \label{tbl:catalog}}
  \tablehead{\colhead{KID} & \colhead{GID} & \colhead{Start} & \colhead{End} & \colhead{Dist} & \colhead{e\_Dist} & \colhead{E\_Dist} & \colhead{Radius} & \colhead{Teff} & \colhead{QFlux} & \colhead{e\_QFlux} & \colhead{Lum} & \colhead{e\_Lum} & \colhead{Ebol} & \colhead{e\_Ebol} & \colhead{E\_Ebol} \\
 &  & GALEX & GALEX & pc & pc & pc & Solar & K & cgs$^\dagger$ & cgs$^\dagger$ & $\mathrm{erg\,s^{-1}}$ & $\mathrm{erg\,s^{-1}}$ & $\mathrm{erg}$ & $\mathrm{erg}$ & $\mathrm{erg}$ \\
 & & Time & Time & & & & radii}
  \startdata
  \hline
9141412 & 3154281294289971861 & 926132291 & 926132401 & 8.8e+02 & 8.6e+02 & 8.9e+02 & 1.06 & 5851 & 3.26e-16 & 9e-18 & 3.16e+31 & 1e+30 & 2.3e+35 & 2.2e+35 & 2.4e+35 \\
9141412 & 3154281294289971861 & 926132353 & 926132401 & 8.8e+02 & 8.6e+02 & 8.9e+02 & 1.06 & 5851 & 3.26e-16 & 9e-18 & 3.16e+31 & 1e+30 & 2.2e+35 & 2.1e+35 & 2.3e+35 \\
9141412 & 3154281294289971861 & 926132470 & 926132574 & 8.8e+02 & 8.6e+02 & 8.9e+02 & 1.06 & 5851 & 3.26e-16 & 9e-18 & 3.16e+31 & 1e+30 & 1.8e+35 & 1.7e+35 & 1.9e+35 \\
9141116 & 3154281294289970910 & 926132590 & 926132690 & 1.4e+03 & 1.4e+03 & 1.5e+03 & 1.31 & 5655 & 7.50e-17 & 6e-18 & 1.84e+31 & 2e+30 & 9.6e+34 & 8.5e+34 & 1.1e+35 \\
9141116 & 3154281294289970910 & 926132732 & 926132795 & 1.4e+03 & 1.4e+03 & 1.5e+03 & 1.31 & 5655 & 7.50e-17 & 6e-18 & 1.84e+31 & 2e+30 & 4.5e+34 & 3.8e+34 & 5.1e+34 \\
5078579 & 3190345275681016588 & 926144930 & 926145034 & 1.7e+03 & 1.6e+03 & 1.8e+03 & 1.46 & 5447 & 7.21e-17 & 8e-18 & 3.97e+31 & 5e+30 & 2.0e+35 & 1.7e+35 & 2.3e+35 \\
5254208 & 3190345275681018094 & 926145151 & 926145246 & 2.9e+03 & 2.6e+03 & 3.3e+03 & 2.57 & 5293 & 8.63e-17 & 8e-18 & 1.68e+32 & 2e+31 & 3.5e+35 & 2.6e+35 & 4.5e+35 \\
8007289 & 3190169353820569784 & 926150174 & 926150265 & 9.2e+02 & 8.1e+02 & 1.1e+03 & -- & 5179 & 4.70e-17 & 5e-18 & 4.74e+30 & 8e+29 & 2.3e+34 & 1.7e+34 & 3.0e+34 \\
8144041 & 3190169353820570884 & 926150362 & 926150441 & 6.1e+02 & 6.0e+02 & 6.2e+02 & 0.70 & 4952 & 5.57e-17 & 6e-18 & 2.27e+30 & 2e+29 & 3.5e+34 & 3.3e+34 & 3.7e+34 \\
8143737 & 3190169353820571145 & 926150622 & 926150784 & 6.2e+02 & 6.0e+02 & 6.3e+02 & 0.72 & 4778 & 1.21e-16 & 5e-18 & 5.37e+30 & 2e+29 & 1.1e+34 & 9.0e+33 & 1.2e+34 \\
  \enddata
\begin{flushleft}
 {\tiny $\dagger$ ~$\mathrm{erg\,\mathring{A}^{-1}\,s^{-1}\,cm^{-2}}$ }
 \end{flushleft} 
\end{deluxetable}

We expect the number distribution of flares with energy to follow a power law: \\
\begin{equation}
\frac{dN}{dE} = k E^{-\alpha}
\end{equation}
where $\frac{dN}{dE}$ is the number of flares occurring per unit energy, $E$ is the flare energy, and  $\alpha$ is a constant, the index of the flare number distribution. In this equation $k$ is a constant of proportionality. Integrating this equation gives a cumulative flare number distribution, \\
\begin{equation}
N(>E) = N_{\rm tot}\left(\frac{E}{E_{\rm min}}\right)^{1-\alpha}
\end{equation}
where $N(>E)$ is the number of flares with energy $>E$, $N_{tot}$ is the total number of flares with energy $>E_{\rm min}$, and $E_{\rm min}$ is a minimum flare energy above which the power-law holds. 
We then turned this number distribution into a flare frequency distribution,
\begin{equation}
FF(E) = N(>E)/N(E_{det} \geq E)/\tau(E)
\end{equation}
where $FF(E)$ is the flare frequency (erg star$^{-1}$ second$^{-1}$) at energy $E$, $N(E_{det} \geq E)$ is the number of flaring stars where flares of energy $E$ can be detected, and $\tau(E)$ is the total observation time for the stars where flares of energy $E$ can be detected. The minimum detectable flare energy for each star was determined by finding the energy of a simple model flare consisting of 3 points above quiescent flux, one elevated 3.5$\sigma$, one 2$\sigma$, and the third 1$\sigma$. While the minimum requirements for a flare per our automated algorithm were two points, one 3.5$\sigma$ above quiescence, with one neighboring point 2$\sigma$ above quiescence, we felt that adding a third elevated point more accurately represented the smallest of our body of flares. We used the maximum likelihood method described by \citet{Clauset2009} to fit $\alpha$ and $E_{min}$ to our data.
We calculate $\alpha$  as\\
\begin{equation}
\alpha = 1 + N_{\rm tot}\left[\sum\limits_{i=1}^{N_{\rm tot}}\ln \left(\frac{E_i}{E_{\rm min}}\right)\right]^{-1} \;\; .
\end{equation}
In this formulation, $\alpha$ depends on $E_{\rm min}$.  Per Clauset et al equation 3.9 we fit $E_{\rm min}$ (and thus $\alpha$) by minimizing the quantity $D$, defined as\\
\begin{equation}
D = \max_{E \geq E_{\rm min}}|S(E) - P(E)|\;\;
\end{equation}
where $S(E)$ is the cumulative distribution function (CDF) of our data, and $P(E)$ is the CDF of the power law model for the given $E_{\rm min}$, \\
\begin{equation}
S(E) = \frac{N(>E)}{N_{\rm tot}N(E_{det} \geq E)\tau(E)}
\end{equation}
\begin{equation}
P(E) = \left(\frac{E}{E_{\rm min}}\right)^{1-\alpha}\frac{1}{N(E_{det} \geq E)\tau(E)}
\end{equation}

\begin{figure}[!htb]
  \includegraphics[width=\textwidth]{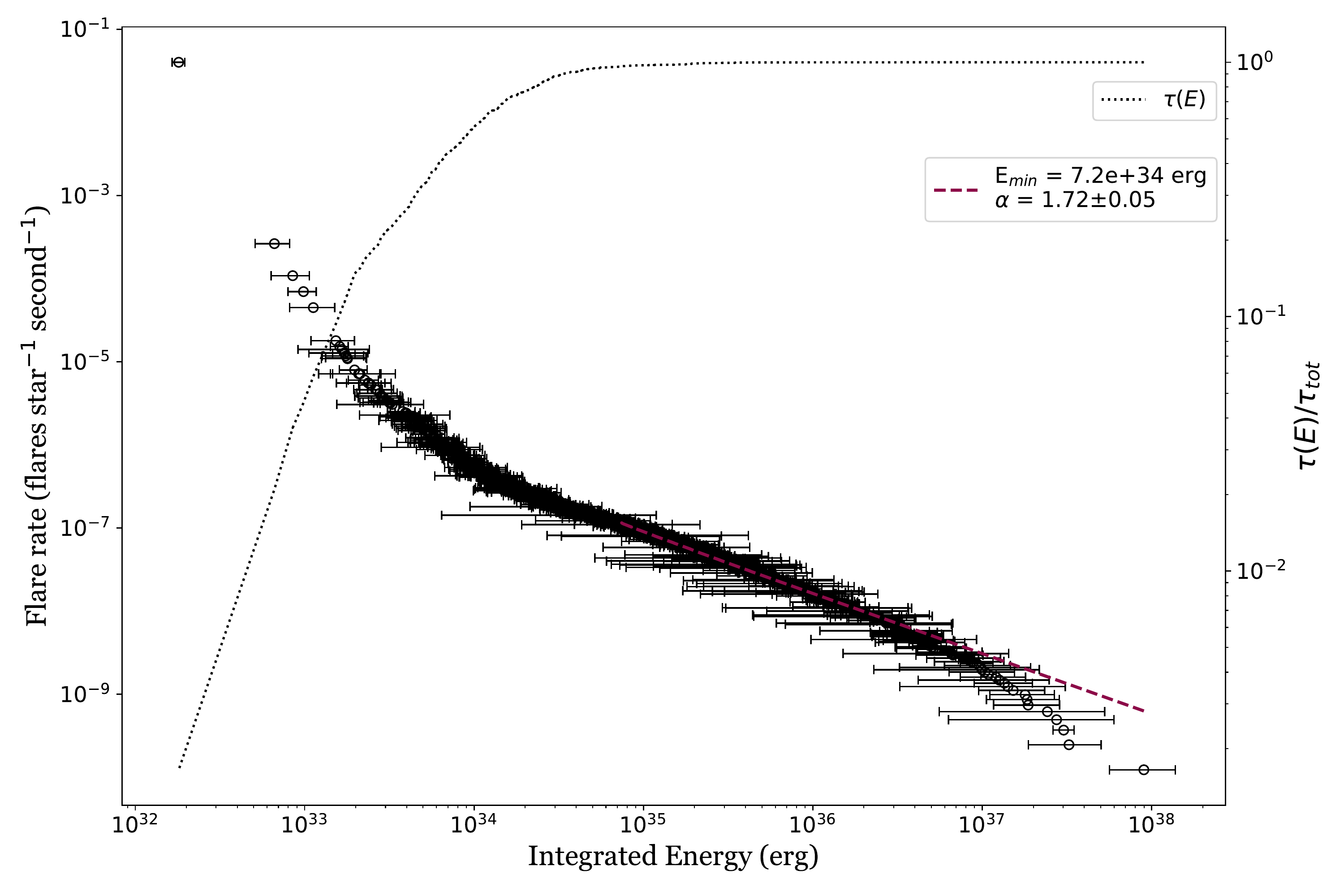}
\caption{Flare frequency as a function of bolometric energy ($E$). Each flare energy has an associated error, which is shown. The red dashed line displays the results of a power-law fit to the distribution; see text for details. The black dashed line is the completeness curve showing percentage of observation time where flares of energy $E$ could have been detected; see text for details.
\label{fig:energyVsNumFlares}
}
\end{figure}

Figure \ref{fig:energyVsNumFlares} shows the flare frequency distribution for our calculated flare energies. For this distribution we considered only flares where the calculated energy was not a minimum value, meaning we discarded all flares that extended past the edge of the observation interval, as well as the flares on stars whose quiescent flux was comparable to the background flux, this left 1705 flares in our sample (see Table~\ref{tbl:dataRedux} step 10). The plot shows a clear power law in the middle section with both ends rolling away from the strict power law. At the highest energies we become observation-limited due to GALEX's visit length; the insensitivity to flares longer than about 30 minutes and expected correlation of flare energy and duration means we are under sampling higher energy flares. At the low energy end the plot turns up sharply, this can be understood by looking at the $\tau(E)/\tau_{tot}$ completeness curve also shown on the plot. We can detect the lower energy flares on far fewer stars, so the flare rate becomes unreliable and artificially elevated.

The error bars reflect the uncertainty in the \citet{BailerJones2018} distance calculations combined with assumed Poisson error on the flux measurements, and errors on the reddening correction. Using \citet{Clauset2009}'s method outlined above, we fit the power law to the cumulative distribution, which gave us an $\alpha$ value of 1.72$\pm$0.05, and $E_{min}$ of $7.20e+34$ ergs. The error on $\alpha$ was determined from the uncertainty in the $\alpha$ fit combined with the uncertainty in the flare energy calculations. As described in \S 2, the distributions for flaring stars do mimic the trends seen for all stars. However, there is a non-uniform distribution of both total observation time and observation interval length, as well as observational sensitivity effect which will be discussed in the next section. 

\subsection{Flare Properties Combined with Stellar Properties}
In this section we combine information derived from the flare properties with stellar properties for a deeper investigation. The stellar properties are discussed in \S3.1 and the flare properties  are discussed in \S3.2.

\begin{figure}[!htb]
  \includegraphics[width=\textwidth]{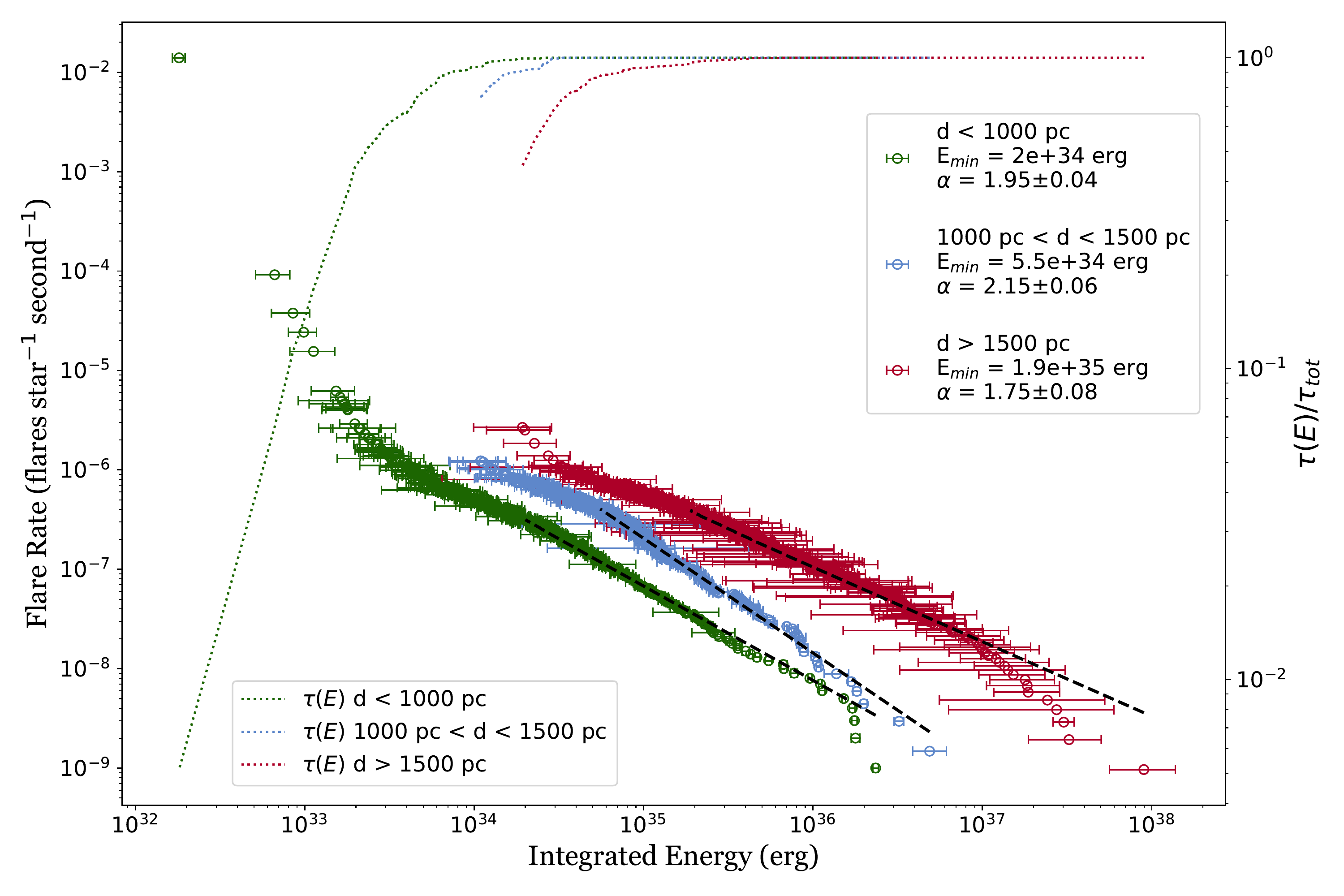}
\caption{Flare frequency as a function of bolometric energy ($E$), partitioned by distance, and associated observation time completeness curves. The distance bins were chosen to contain approximately equal numbers of flares. The influence of selection effects, in seeing higher energy flares on more distant stars, is apparent, while the shape of the distribution is approximately the same.
\label{fig:energyByDistance}
}
\end{figure}

Because of the large span of distances in our sample, we investigated how the sensitivity limitations affect the aggregate flare statistics. As mentioned above, the sample of flaring stars for which distance measurements are available is very large, 999 out of the 1,021 that passed the flare filtering. Figure \ref{fig:energyByDistance} shows the flare frequency distribution for our calculated flare energies, split up by distance. The distance splits were chosen so that roughly a third of our flare population fell in each bin. Each frequency distribution shows clear evidence of a power law with departures for both high and low energy flares. The $\alpha$ values are similar but not the same for each bin (see Table \ref{tbl:fits}). Because the flare energy determination increases as $d^{2}$, there is an expectation that the most distant stars will tend to have the most energetic flares, and that is borne out by the  three distributions containing progressively more energetic flares, as well as having progressively higher $E_{min}$ values fit to the distributions.

\begin{deluxetable}{lrr}
  \tablewidth{0pt}
  \tablecolumns{4}
  \tablecaption{Cumulative energy distribution fits
    \label{tbl:fits}}
  \tablehead{\colhead{Criteria} & \colhead{$E_{min} (10^{34}erg)$} & \colhead{$\alpha$}}
  \startdata
  \hline
  \multicolumn{3}{c}{All flares} \\
  \hline
   & 7.20 & $1.72\pm0.05$ \\
   \hline 
   \multicolumn{3}{c}{Partitioned by distance}\\
   \hline
   $d < 1000$ pc & 2.00 & $1.95\pm0.04$ \\
   $1000 < d < 1500$ pc & 5.53 & $2.15\pm0.06$ \\
   $d>1500$ pc & 18.91 & $1.75\pm0.08$\\
   \hline 
   \multicolumn{3}{c}{Partitioned by $T_{eff}$}\\
   \hline
   T $< 5550$ K & 5.23 & $1.49\pm0.06$ \\
   T $> 5550$ K & 6.40 & $1.80\pm0.06$
  \enddata
\end{deluxetable}

\begin{figure}[!htb]
  \includegraphics[width=\textwidth]{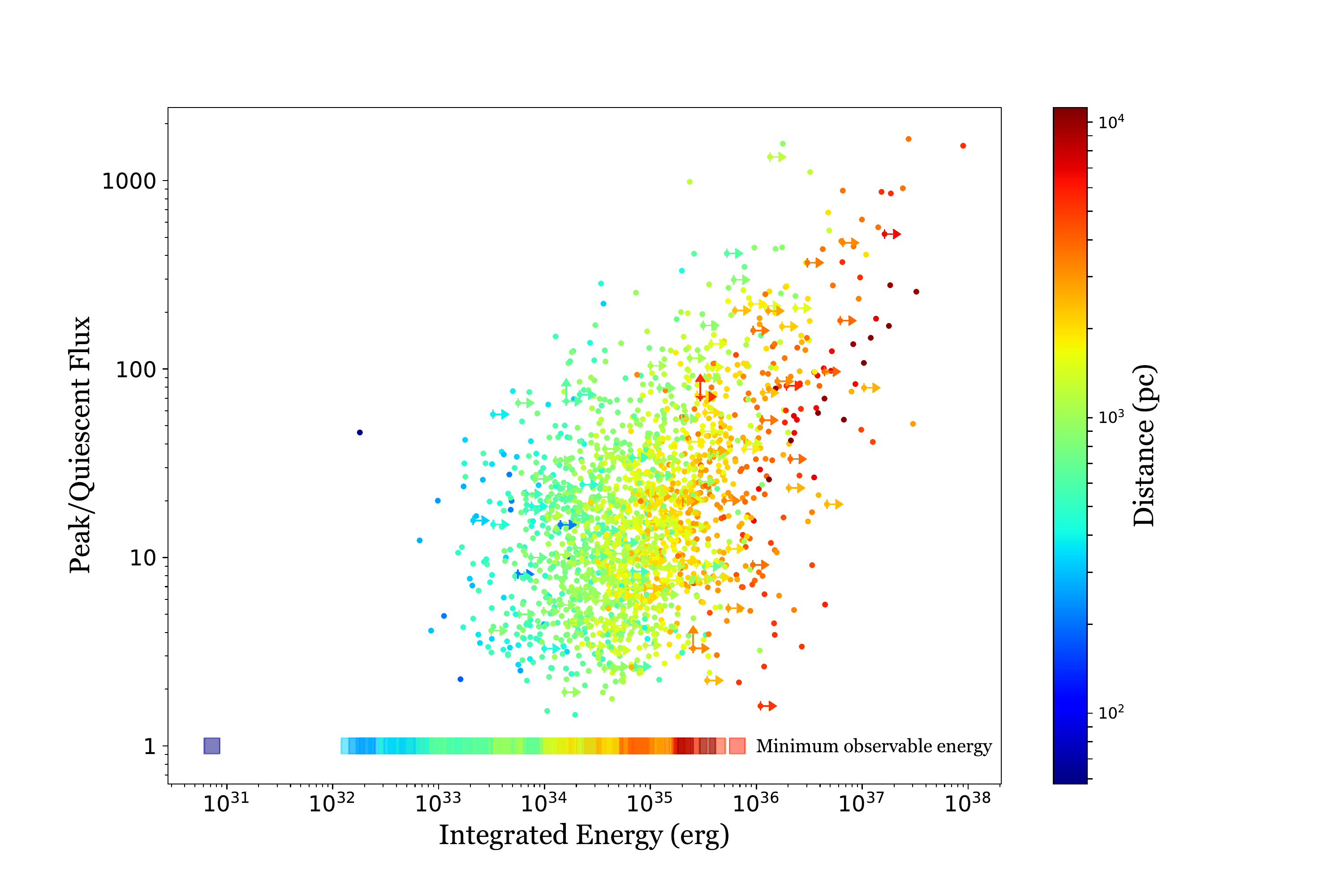}
\caption{Plot of peak flare flux vs. integrated flare energy, with points color coded by distance. The bar across the bottom shows the minimum observable flare energies, also color coded by distance. Arrows indicate when the calculated point is a minimum either due to the flare extending past the observation interval, or the quiescent flux being comparable to the background flux. This plot illustrates the relation between flare energy and distance. It also reveals that while the lower energy flares have similarly small peak flux enhancements, higher energy flares have a wide range of peak flux enhancement. 
\label{fig:ePeakDist}
}
\end{figure}

\begin{figure}[!htb]
  \includegraphics[width=0.8\textwidth]{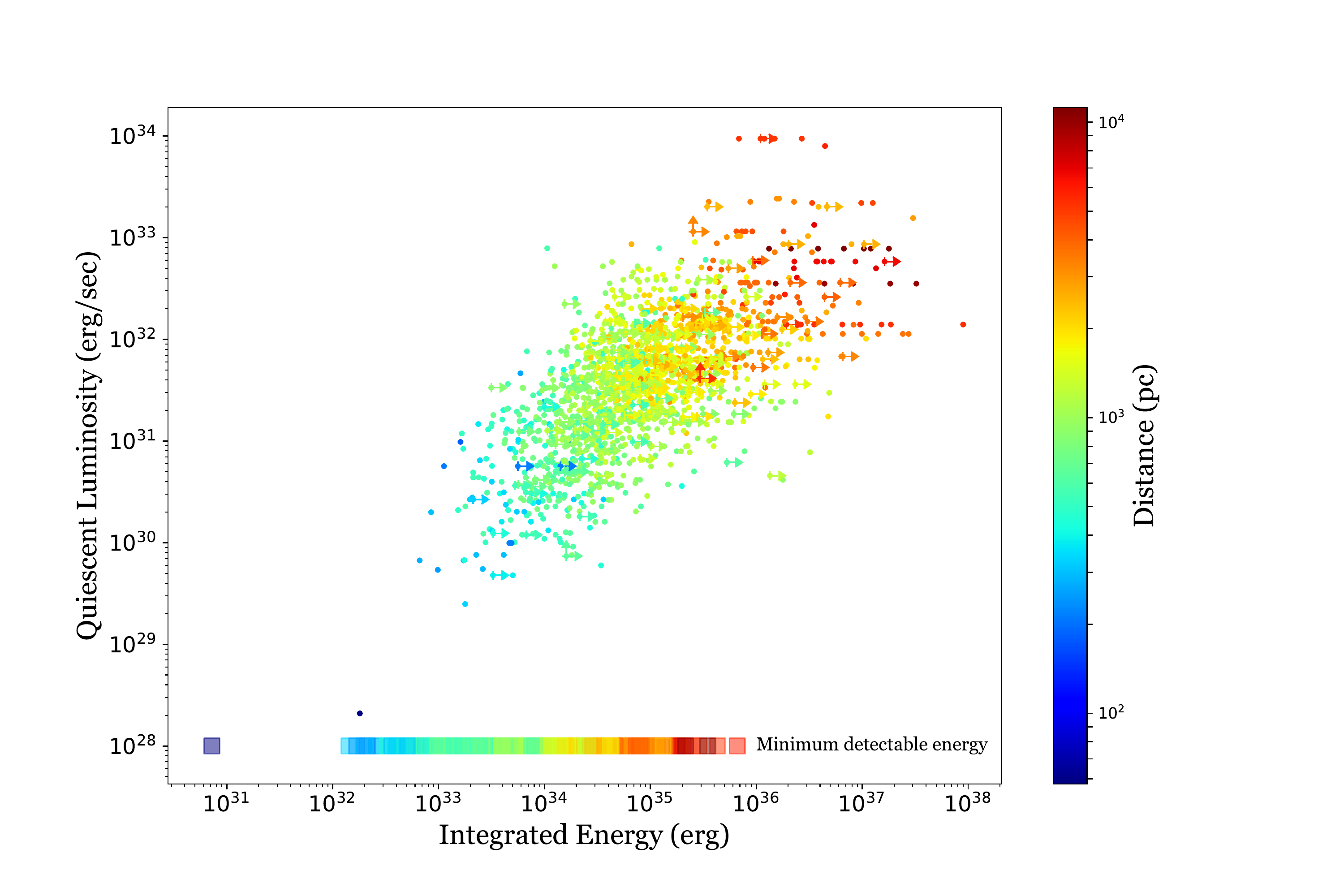}
  \includegraphics[width=0.8\textwidth]{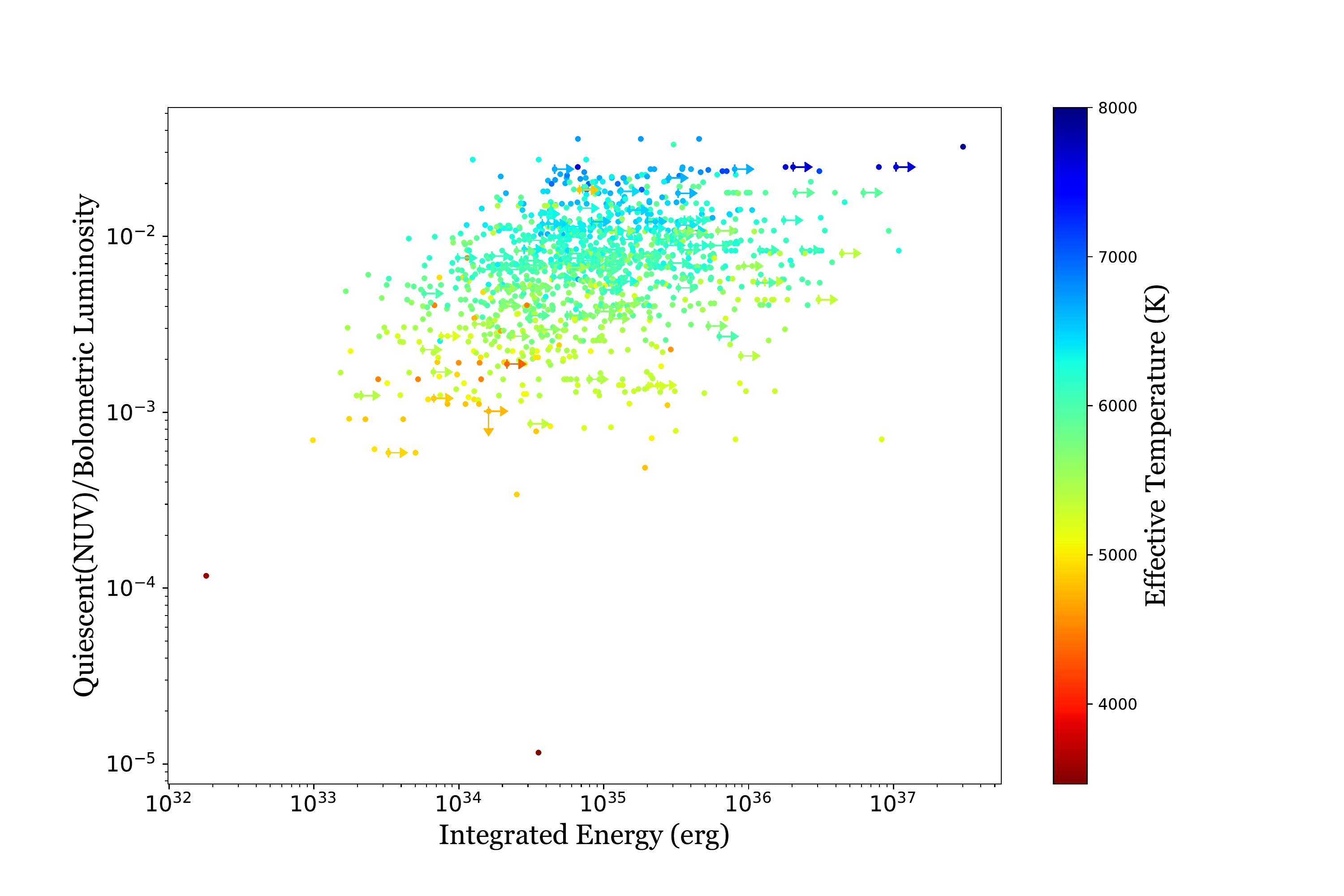}
\caption{Plot of stellar quiescent NUV luminosity vs. integrated flare energy, with points color coded by distance (top), and effective temperature (bottom). Minimum observable flare energies colorized by distance are shown across the bottom of the top plot. Arrows indicate when the calculated point is a minimum either due to the flare extending past the observation interval, or the quiescent flux being comparable to the background flux. There is a strong correlation between  quiescent luminosity and flare energy, as both have the same dependence on distance. The result is that more distant stars have both higher quiescent luminosity and more energetic flares. 
\label{fig:eLumDist}
}
\end{figure}

We investigated further the impact of stellar distance on the flare quantities we derived. Figure~\ref{fig:ePeakDist} displays a scatter plot of flare peak enhancement as a function of the integrated energy, with points sorted by stellar distance, from red (most distant) to blue (closest). The minimum observable flare energy for each star is shown in the bar across the bottom of the plot, also colorized by distance. The ``rainbow'' proceeding from right to left in the plot is indicative of distance impacting energies, as expected.  Distance does not appear to affect the peak flare enhancement however; Figure~\ref{fig:flareDurPeakFlux} shows a scattering of flare durations for the span of peak flux enhancements. The smallest energy flares do tend to have a restricted range of flux enhancements, which lie to the short end of the flux enhancement distribution, but this broadens as the flare energy increases. Figure~\ref{fig:eLumDist} \textit{top} plots the quiescent NUV luminosity of the star against the integrated flare energy. The former quantity was derived by converting the  quiescent flux measured in \S 2.1 using the stellar distance. The points are colorized by the stellar distance. Here there is a clear correlation between these two quantities, which both have the same dependence on distance.

\begin{figure}
    \centering
    \includegraphics[width=\textwidth]{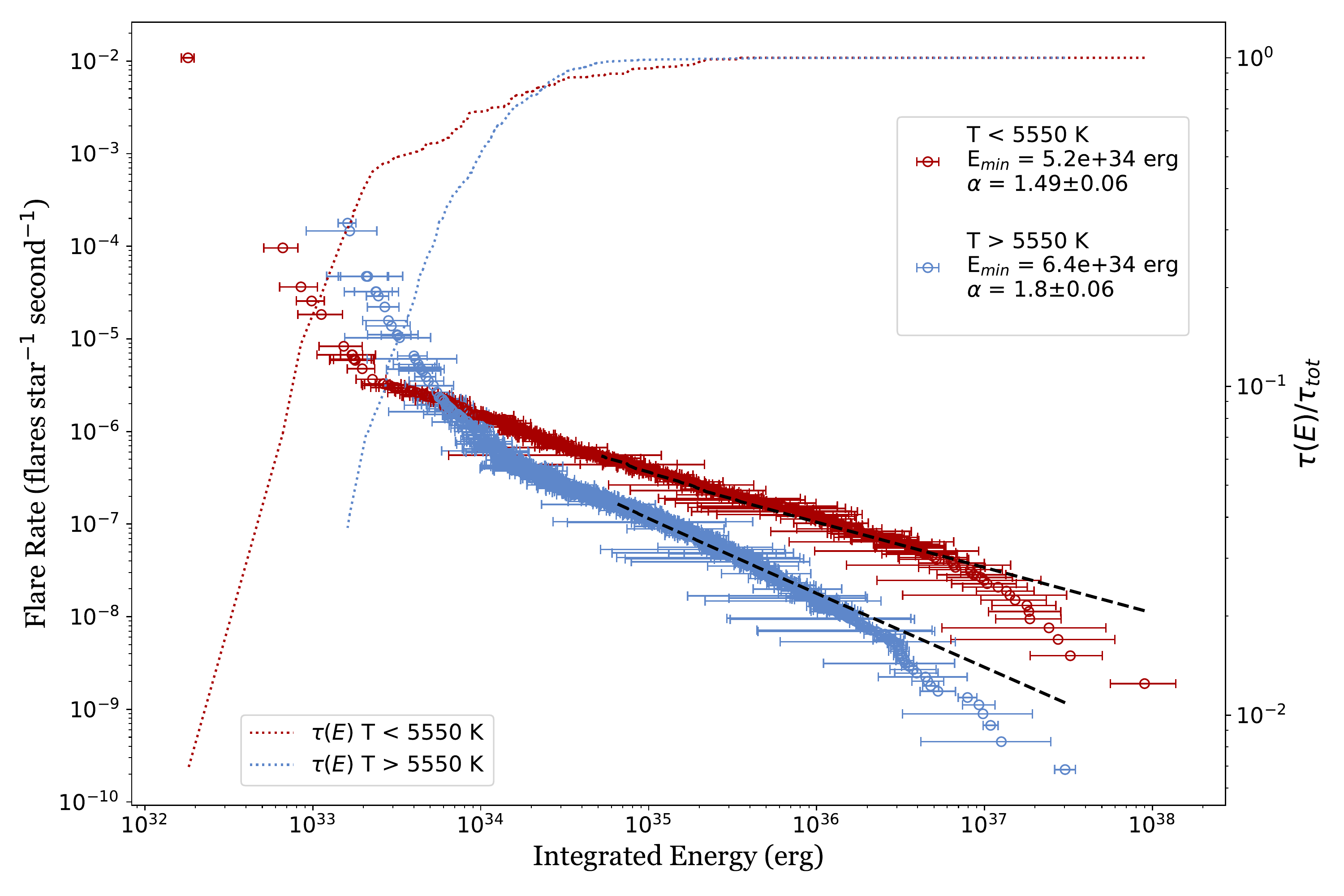}
    \caption{Flare frequency as a function of bolometric energy ($E$), partitioned by T$_{\rm eff}$, and associated observation time completeness curves.}
    \label{fig:energyByTeff}
\end{figure}

We also explored the impact of effective temperature on the cumulative energy distribution, splitting the energy distribution at $T_{eff} = 5550 K$ (Figure \ref{fig:energyByTeff}) where a dip in the $T_{\rm eff}$ distribution occurs (see Figure \ref{fig:teff_rad_dist} inset). The $\alpha$ values are listed in Table~\ref{tbl:alphas}. The calculated $\alpha$ values differ, with higher temperature stars having a steeper slope to the cumulative energy distribution, however we must be careful not to read too much into this as there are significantly more flares on higher temperature stars (641 vs. 1244).  The bottom panel of Figure~\ref{fig:eLumDist} shows the dependence of quiescent NUV luminosity ($L_{q}$) on flare energy ($E_{\rm fl}$), normalized to bolometric luminosity, with $T_{\rm eff}$ as the color. In addition to the distance influence noted above, there is also an effective temperature gradient. The coolest stars have the smallest $L_{q}$ and $E_{\rm fl}$, progressing gradually to higher values of $L_{q}$ and $E_{\rm fl}$ for the hotter stars. This does not hold universally, though, as some of the stars with the highest $L_{q}$ and $E_{\rm fl}$ have $T_{\rm eff}\sim$5000 K.

\begin{figure}[!htb]
  \includegraphics[width=\textwidth]{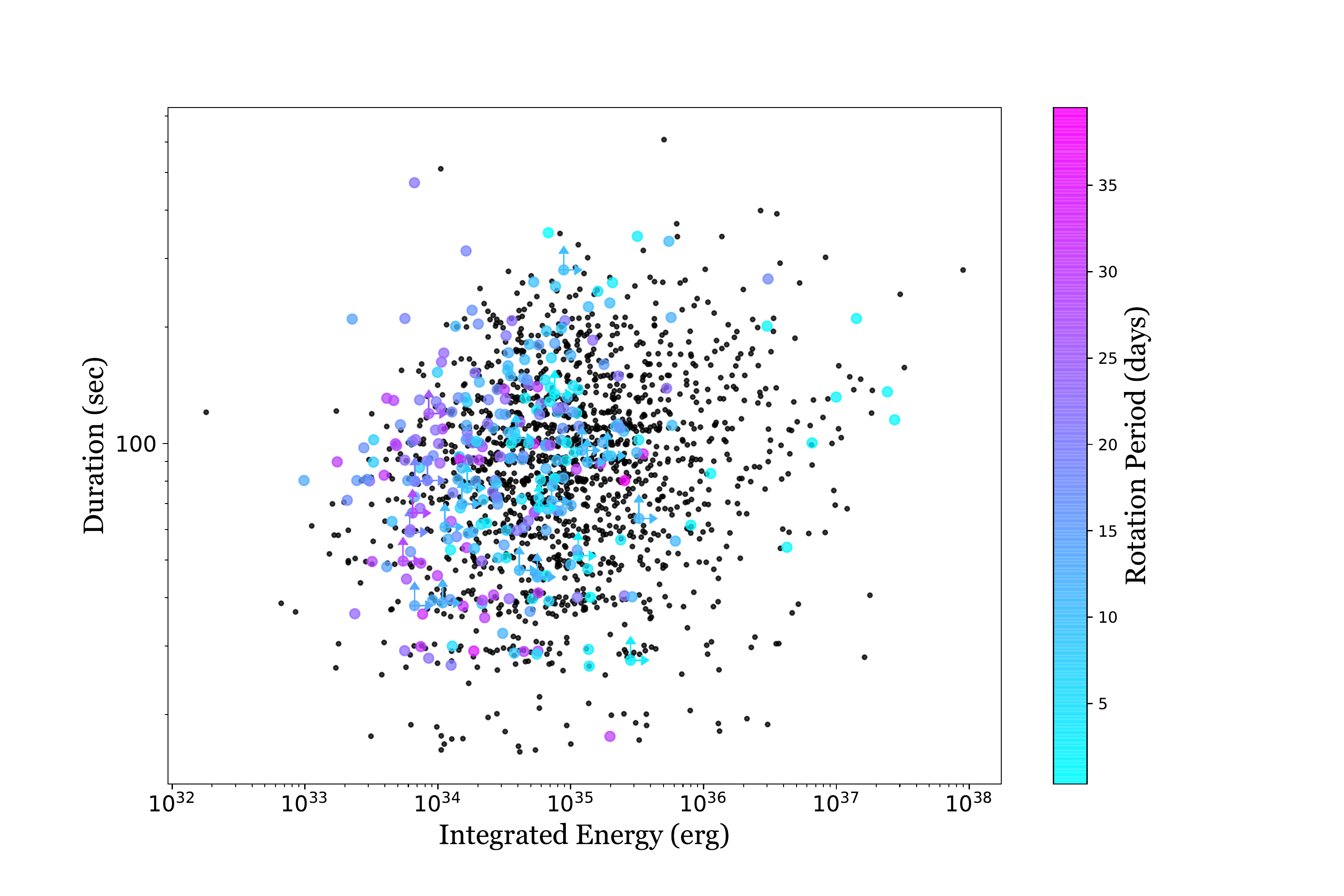}
\caption{Scatter plot of bolometric radiated flare energy against duration, with a color coding for stars which have defined periods (stars without defined periods are left black). Arrows indicate when the calculated point is a minimum either due to the flare extending past the observation interval, or the quiescent flux being comparable to the background flux. There is not a strong correlation between flare energy and duration. Higher energy flares tend to be on stars with shorter rotation periods, although the two stars with the longest rotation periods have flare energies inconsistent with this expectation.
\label{fig:eDurRot}
}
\end{figure}

Figure~\ref{fig:eDurRot} plots the duration of the flares against the flare energies, and uses a color coding for those stars which have  good-quality rotation period measurements (see \S 3.1 for details). While there is a general trend for higher energy flares to be found on stars with shorter rotation periods, the two stars with the longest rotation periods have flare energies of several times 10$^{35}$ erg, roughly in the middle of the range of flare energies.

\begin{figure}[!htb]
  \includegraphics[width=\textwidth]{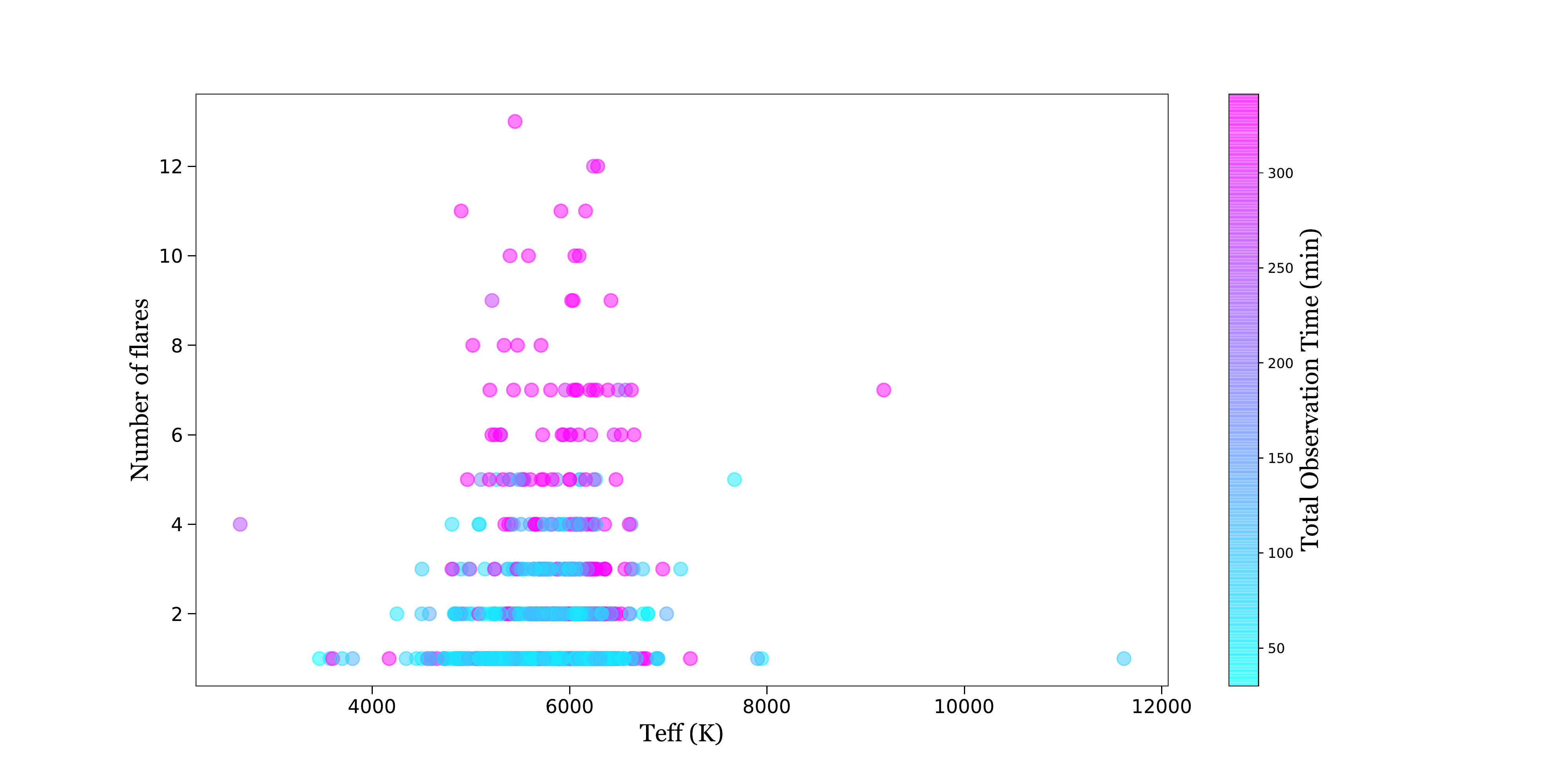}
\caption{Plot of the number of observed flares vs. effective temperature, color coded by total observation time.  While the stars with multiple flares were observed for longer periods of time, many of the stars with longer observation times only exhibited one or a few flares.
\label{fig:teffNumObsTime}
}
\end{figure}

Figure~\ref{fig:teffNumObsTime} plots the number of detected flares vs. the effective temperature of the flaring stars colored by total observation time. There is a clear correlation between total observation time and number of detected flares, due to there being more time to potentially detect flares, however there is not a similar correlation between effective temperature and number of flares.

\section{Discussion}
The wavelength region used in this study of stellar flares, combined with the nature of the stars probed for their flares, makes for a unique combination which has not been studied to date. Here we discuss the results, broken into the types of flaring stars and the characteristics of the flares.

\subsection{Types of Flaring Stars}
The characteristics of these stars reflect the selection used for the Kepler mission.  While there are some flaring stars off the locus of main-sequence stars shown in Figure~\ref{fig:colorMagnitude}, these are a minority of stars. As described in \S 3.1, the stars studied for flares are restricted to being near the main sequence.  This study is able to take advantage of the additional information about the properties of stars targeted in the main Kepler mission to interpret our results in addition to independent information about magnetic activity from the data itself. The flaring stars range in T$_{\rm eff}$ from about 3000 to 11,000 K, and the bulk are between 4500-6500 K. The flaring stars range in radius from 0.5-15 R$_{\odot}$, with the bulk being between 1-2 R$_{\odot}$. The approximate spectral types of most of these stars are then $\sim$mid-F to mid-K dwarfs. The distances to the stars span quite a large range, from 57 pc to 10 kpc. 

The datasets analyzed in this paper contain additional information about magnetic activity beyond the detection of flares, which is the main focus of the present paper. The quiescent NUV luminosity is another indicator of nonradiative heating caused by the presence of magnetic fields in stellar outer atmospheres. \citet{linsky2017}'s recent review summarizes characteristics of stellar chromospheres. The emission contribution from the stellar photosphere decreases at increasingly shorter wavelengths, so that by $\sim$2500 \AA\ it is necessary to include emission from hotter chromospheric material with elevated pressures above photospheric level to account for all of the observed spectrum. The quiescent NUV spectrum is dominated by lines of Mg~II and a forest of Fe~II lines, in addition to weaker continuum edges \citep[][and references therein]{linsky2017}. We do not have spectrally resolved emission to subtract the photospheric blackbody contribution from our integrated quiescent NUV luminosities, and it is beyond the scope of the present paper to do so. Instead we use the measure of quiescent NUV luminosity as an indicator of magnetic activity. 

Stellar flares are an indicator of transient magnetic activity. Thus a naive expectation is that trends established for other magnetic activity indicators should also hold for flare properties. There are well-established trends between other magnetic activity indicators and rotation that apply to cool main sequence stars \citep[e.g.,][]{skumanich1972}. The most obvious one is at the heart of the gyrochronology technique \citep{barnes2003}, in which magnetic activity decays with age as a single star loses angular momentum and slows its rotation period; the measurement of magnetic activity is then used to gauge stellar age. More recently, \citet{davenport2019} examined the evolution of flare activity with stellar age, reproducing the activity-age relationship expected from the more fundamental activity-rotation correlation.  There is no evidence in our data for a rotation period dependence on stellar flare properties like flare duration or integrated flare energy, although the small number of stars in the present sample with bona fide rotation periods prevents a deeper investigation.

The rotation-activity correlation extends to binary systems. Stars in a binary system  can rotate rapidly because of coupling between spin and orbital angular momentum, and also experience enhanced levels of magnetic activity \citep{basri1987} resulting from their rapid rotation. The extent to which binarity may affect the activity level is not a settled subject matter: while \citet{basri1985}, \citet{basri1987}, and \citet{stern1995} found little to no evidence of enhanced activity measures for binaries as compared to single stars with the same rotation period, \citet{morgan2016} determined that non-interacting close binaries composed of white dwarf-M dwarf stars with separation $<$10 a.u.  had a higher flare rate than single M dwarfs. Only 14\% of our sample have defined periods in \cite{McQuillanEtAl2014}, which range from less than one day to just under 40 days. The binary rate of other studies utilizing data from the main Kepler mission for solar-like stars can be high, 30-50\% (D. Nogami, private communication). As we do not have constraints on the binarity of these stars we cannot determine whether the presence of a companion affects the magnetic activity signatures being diagnosed. As Figure~\ref{fig:eDurRot} shows, the most energetic flares in our sample do seem to originate from more rapidly rotating stars, but the slowest rotating stars have flare energies of several times 10$^{35}$ erg, much larger than the lowest flares seen.

Our data sample suffers from several biases, the most fundamental of which is a sensitivity limitation due to distance. This is most clearly conveyed in Figures~\ref{fig:energyByDistance} through \ref{fig:eLumDist}. The most distant stars do have the highest energy flares, but they do not necessarily have the largest flare enhancements. Figure \ref{fig:ePeakDist} shows that while the lower energy flares tend to have correspondingly small peak flux enhancements, the higher energy flare peak flux enhancements span the entire range of the dataset. The sample contains both low-energy, low-enhancement flares on nearby stars, as well as a spread of peak flare fluxes on the most distant stars. There is a strong relationship between integrated flare energy, quiescent luminosity, and distance (Figure \ref{fig:eLumDist}). Here the effect of  observational sensitivity is most evident, with more distant stars having larger quiescent NUV luminosities as well as larger flare energies. The left side of the plot has a sharp edge to the distribution of points, which shows the sensitivity limits to lower energy flares. On the right side there is a horizontal spread of flare energies at a given flare peak flux, which indicates less of a sensitivity limit to detecting larger flares. The minimum observable flare energies, shown across the bottom in the top plot backs this up as the distance colors generally align with the left cutoff of the spread, with the observed flares spread out to the right.

As discussed in \S3.2.3, the heterogeneous nature of the data collection, and the distribution of total observation time per star (see Figure~\ref{fig:obsHist}) make a determination of the frequency occurrence of flares more challenging.  In our treatment we calculated the flare rate by starting with the cumulative number distribution and then, for each energy dividing by the number of stars (from our sample of flaring stars) where a flare of that energy was observable, and the total observation time for all stars where a flare of that energy could be observed. This is not a perfect metric, as it does not take into account the differences in observation interval length, however as shown in Figure \ref{fig:flareDurIntLength} the duration of the vast majority of our flares are less than half the length of the observation interval, indicating that most of our flares are not affected by the observation interval. Figure~\ref{fig:teffNumObsTime} shows that there is a general trend to see more flares on stars for a longer total observation time, although there are exceptions. There does not appear to be a skewing of this trend toward stars of differing T$_{\rm eff}$, as one might expect if low-mass stars preferentially produced more high energy flares than more solar-like ones. In fact, Figure~\ref{fig:energyByTeff} shows a similar relative distribution of flare energies for stars hotter and cooler than a T$_{\rm eff}$ cut at 5550 K. 

The other obvious bias in our sample relates to the quiescent amount of NUV emission. This is determined from the nonflaring portions of the NUV light curve. As demonstrated in Figure~\ref{fig:eLumDist}, there is a strong correlation between the time-averaged measure of stellar magnetic activity, the quiescent NUV luminosity, and the measure of the transient stellar magnetic activity, the integrated flare energy. The relationship between quiescent luminosity and flare energy is expected, because both are indicators of magnetic activity. 

\citet{Davenport2016} produced a catalog of Kepler flares from a search on every available Kepler light curve from the main mission, with a catalog size of 4,041 flaring stars. Somewhat surprisingly, only 13 stars are in common between these two studies. There are a number of possible reasons for the small size of this overlap. The first originates from the different time regimes in which the two missions obtained data; the present study has time sampling of 10 seconds, but data accumulated only over a maximum of several hours per star. The Kepler mission, in contrast, obtained mostly 30-minute cadence data over timescales of months to years. The flares found in the present study have vastly shorter durations; the range found here spans under a minute to about 10 minutes, while \cite{Davenport2016}'s flares are much longer, with no information on specific flare durations. Other studies of flares on solar-like stars observed with Kepler \citep{maehara2012,nogami2014,maehara2015} span the range 20--300 minutes. The second reason stems from the stringent conditions imposed by \cite{Davenport2016} on the number of putative flares that must be detected on a given star before it merited a place in that catalog. Due to the number of light curves and flare detections, \cite{Davenport2016}'s sample could not be manually validated, which required imposing a threshold of at least 100 total candidate flare events  per star, with at least ten events having energies greater than the local 68\% completeness threshold. Because the total observation time and number of candidate flares in the present study is much smaller, manual validation of all of the candidate flares can be accomplished easily. Indeed, as Figure~\ref{fig:teffNumObsTime} demonstrates, the maximum number of flares per star in the present sample is only 13. This opens up the possibility to validate some of the flare candidate stars from \cite{Davenport2016}'s sample that did not meet the threshold by adding in the use of multi-wavelength flare observations.

Figure \ref{fig:eDurRot} reveals a lack of a strong correlation between energy and flare duration. There is a noticeable correlation between rotation period and flare energy, with larger flares tending to be on faster rotating stars. However, some of the slowest rotating stars exhibit flare energies  in the middle of this range, so the dependence does not appear to be strong. This overall trend fits the rotation-age-activity relation in single stars, with more active stars having larger flares, and thus being assumed to be younger stars. We do see flares even on slowly rotating stars however, which indicates that we are seeing flares on even relatively ``inactive'' stars. See \S 4.1 for a discussion of binary contamination in our sample.

A natural question to consider is to what extent the present results can reveal information about the maximum flare energy the Sun might produce. This question is relevant to a fuller understanding of the behavior of our Sun, as well as extrapolating to the conditions which might be present on planets around other solar-like stars. \citet{maehara2012} established the existence of energetic flares on solar-like stars found in the Kepler bandpass, estimating the occurrence rate of a flare on a solar-like star (5600 $<$T$_{\rm eff}<$6000K, $P_{\rm orb} >$10 days) with energy in excess of 10$^{34}$ erg in the Kepler bandpass happening once every 800 years. The present data illustrate that solar-like stars can release flares with energies of up 10$^{38}$ erg, or about two orders of magnitude larger than what was constrained by \citet{maehara2012} and subsequent work. We can use the additional information in our dataset on magnetic activity to constrain this further. Figure~\ref{fig:eLumDist} plots the quiescent stellar NUV luminosity against flare energy for our sample. \citet{fontenla2015} modelled the solar spectral irradiance in the NUV, finding good agreement with observations. They determined the 2000-3000 \AA\ irradiance to be 14.5 W m$^{-2}$, or an NUV luminosity of approximately 4$\times$10$^{31}$ erg s$^{-1}$. This wavelength range is approximately the GALEX NUV bandpass, and we use this as a gauge to separate stars which have a more solar-like magnetic activity level.If we apply this cutoff to Figure~\ref{fig:eLumDist} we still find stars with flare energies ranging up to 10$^{36}$ erg. The Sun's NUV luminosity can vary by large factors; \citet{fontenla2015}'s Fig.~12 illustrates wavelength-dependent relative changes of 2--40\% due to solar surface activity, and \citet{lean2000} quantified irradiance changes in this wavelength region between 0.2 and 10\%. If we filter the stars in the present sample for the range 5600$<T_{\rm eff}<$6000 K, 0.5R$_{\odot}< R_{\star} < $1.5 R$_{\odot}$, 20 $<P_{\rm rot}<$ 40 days, and L$_{q, NUV} <$ 6$\times$10$^{31}$ erg s$^{-1}$, we find a total of 19 flares on 13 stars, with energy ranges spanning 5$\times$10$^{33}$-9$\times$10$^{34}$ erg. While still much larger than the energetic releases seen on the Sun, it is a much more restricted range than the full dataset considered here. \citet{notsu2019} recently reported on a restricted sample of highly vetted solar-like stars found to go undergo superflares in the Kepler data, and determined for slowly rotating solar-like stars the maximum flare energy is less than 5$\times$10$^{34}$ erg, in agreement with our results.

\begin{deluxetable}{lcrcrr}
\rotate
  \tablewidth{0pt}
  \tablecolumns{6}
  \tablecaption{Comparing flare frequency distribution power laws across datasets. We can see that the calculated $\alpha$ values generally agree with each other, as they range from 1.5 to 2.2. Most of the studies are in the optical range with a few in the x-ray and ultraviolet. The results from the two solar studies included aline well with the stellar studies.
    \label{tbl:alphas}}
  \tablehead{\colhead{Reference} & \colhead{Stellar Type} & \colhead{Num. Flares} & \colhead{Bandpass} & \colhead{$\alpha$} & \colhead{Err.}}
  \startdata
  \hline
   This paper & Kepler & 1,705 & NUV & 1.72 & 0.05 \\
   \citet{maehara2015} & Kepler & 187 & Kepler & 1.5 & 0.1 \\
   \citet{Shibayama_etAl_2013} & G-type & 1,547 & Kepler & 2.2 & \\
   \citet{Veronig_etAl_2002} & Solar & 50,000 & soft x-ray & 1.88 & 0.11 \\
   \citet{Wu_Ip_Huang_2015} & G-type & 4,944 & Kepler & 2.04 & 0.17 \\
   \citet{Wu_Ip_Huang_2015} & G-type (freq flarers) & 1,336 & Kepler & 1.81 & 0.03 \\
   \citet{Aschwanden_EtAl_2000} & Solar & 81 & EUV &  1.79 & 0.08 \\
   \citet{wolketal2005} & Young Sun & 41 & X-ray & 1.7 & \\
   \citet{Davenport_etAl_2016} & M dwarf & 66 & Optical & 1.68 & 0.10\\
   \citet{Howard_etAl_2018} & M dwarf & 24 & Optical & 1.98 & -0.02/+0.24 \\
   \citet{Loyd_EtAl_2018} & M dwarfs & 80 & FUV & 1.76 & 0.1 \\
   \citet{Audard_etAl_1999} & Solar analog & 28 & Coronal & 2.2 & 0.2 \\
   \citet{Paudel_etAl_2018} & Ultracool dwarfs & 283 & Kepler & 1.52-2.03 & \\
   \citet{Hawley2014} & M dwarfs & 1305 & Kepler & 1.52-2.32 & \\
  \enddata
\end{deluxetable}

\subsection{Characteristics of Stellar Flares}

\subsubsection{Flare Spectral Energy Distribution}
The GALEX NUV observations span a wide wavelength range, from 1771-2831 \AA. While these data provide no spectral information within that bandpass, we rely on solar and stellar flare studies to inform the likely contributors to the flare flux. The NUV spectral region has not had as many observational constraints as the far UV region in flare studies. Few NUV stellar flare spectra exist at all, and the few that do were obtained either for solar flares or on nearby M dwarfs. Flares observed in the UV are often associated with the more impulsive phases of solar flares, starting with early observations showing a  close temporal association between UV and hard X-ray emission \citep{cheng1981}.   \citet{welsh2006} reported on high time resolution NUV+FUV flares seen with GALEX on nearby M dwarfs, and \citet{hawley2007} presented high spectral resolution NUV flare measurements of an M dwarf flare.  From these two studies the contribution of emission lines relative to continuum emission could be determined; the main emission lines in the flare NUV spectrum were Mg~II, Fe~II, Al~III, and C~III.  However the main emission component overall was a continuum component.  Recent results from solar flares observed from space \citep{heinzelkleint2014,kleint2016} demonstrate an NUV spectrum originating largely from Hydrogen Balmer continuum emission. The formation of the NUV emission appears to  originate from an impulsive thermal and nonthermal ionization caused by the precipitation of electron beams through the chromosphere. This explains the temporal correlation with solar flare hard X-ray emission observed previously.  More recently \citet{kowalski2018} presented accurately calibrated NUV flare spectra at high time cadence on an M dwarf, again finding a large flux enhancement due to continuum radiation. They commented that the oft-used 9000 K black-body used to describe blue-optical stellar flare emission \citep{Hawley2003} under-predicts the NUV continuum flare flux by a factor of two. Based on general similarities in radiative properties between solar and stellar flares studied thus far \citep{ostenbook}, it is likely that a combination of line and continuum emission enhancements are present in the NUV flare flux from the flares being considered, but we cannot speculate about the relative contribution of one versus the other. These sources originate from different layers of the stellar atmosphere: singly- and doubly-ionized emission lines likely originate in the chromosphere and the Balmer continuum emission also originates from  the chromosphere. Some lines such as Mg~II exhibit absorption components and self-reversals indicating optical depth effects in the atmosphere, while other lines such as Fe~II appear to be optically thin. Any hot black-body emission might originate further in the photosphere. 
\subsubsection{Flare Frequency Distribution}

In addition to calculating the power-law index for the flare frequency of all the flares taken together, we split them up by distance and temperature. Despite small differences in $\alpha$ values when splitting the flare population in different ways it remains restricted, as shown in Table \ref{tbl:fits} all our calculated $\alpha$ values fall between 1.49 and 2.15. With our error bars this range agrees with the results of other flare frequency studies as shown in Table \ref{tbl:alphas}, where the calculated $\alpha$ range is 1.52-2.32. The $\alpha$ value for low temperature stars only ($T < 5550$) is slightly lower ($\alpha = 1.49 \pm 0.06$) than any of the values shown in Table \ref{tbl:fits}, however the associated error bar does overlap, and this is the distribution including the smallest number of flares (461).  It is important to note that these values come from both solar and stellar flare studies, and across wavelengths ranging from optical to x-ray. This suggests that the same physical processes are at work in both solar and stellar flares and across wavelengths.

\subsubsection{Timescales}
\begin{figure}[!htb]
  \includegraphics[width=\textwidth]{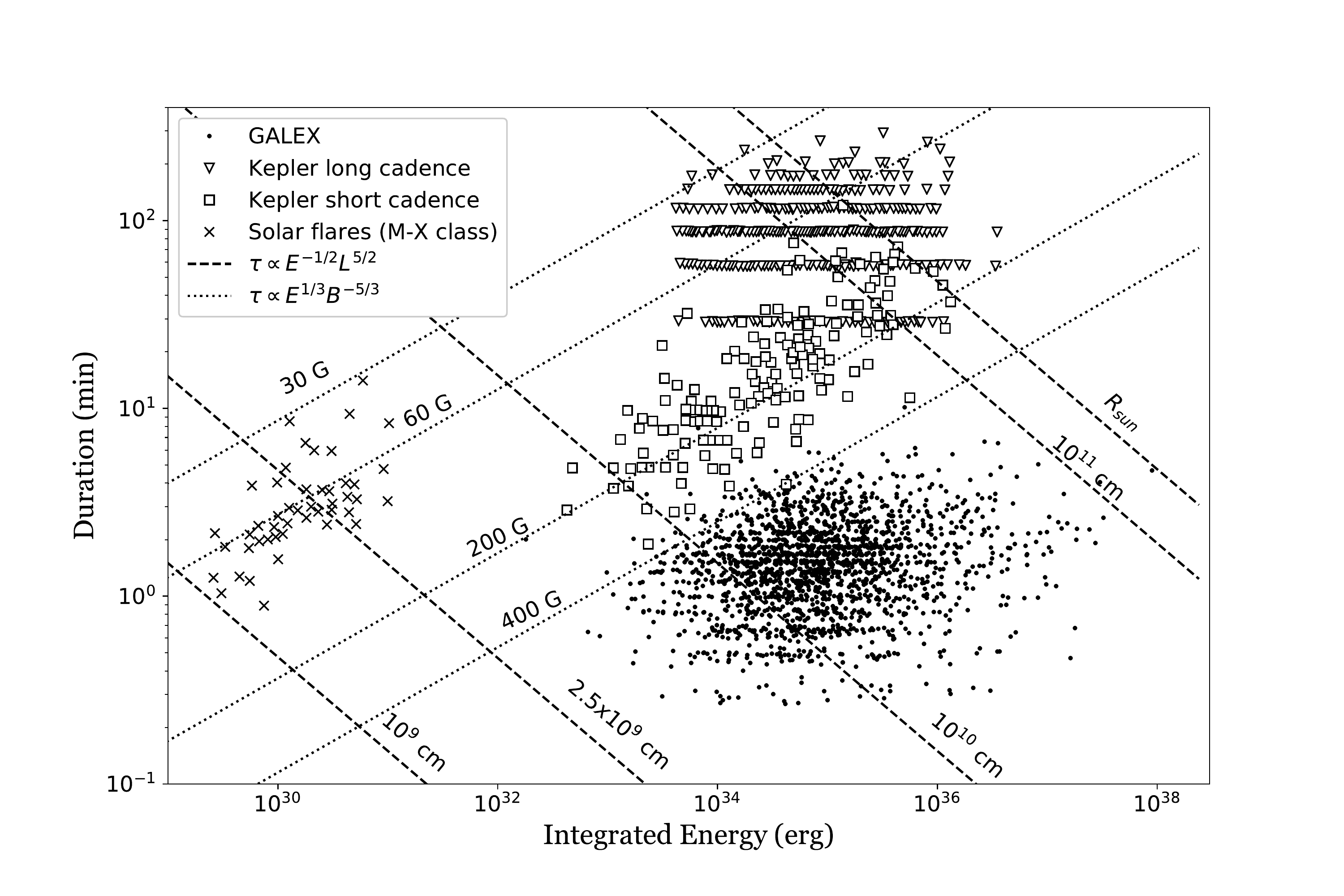}
\caption{Expanded plot of flare duration and flare
energy, including white light solar and stellar flare datasets from \citet{namekata2018}.  The dotted and dashed lines show the theoretical scaling laws derived in \citet{namekata2018}, where B is the magnetic field strength in the flaring region and L is flare length scale. The flares in our sample originate in active regions with $B>$400 G and length scales near 10$^{10}$ cm; see text for details.
\label{fig:namekata}
}
\end{figure}

Following the announcements of stellar superflares seen by Kepler at optical wavelengths \citep{maehara2012}, there has been much progress on advancing understanding of white-light flaring on G stars observed by Kepler. While flares occur more frequently on faster-rotating (and therefore likely younger) G stars, \citet{nogami2014} found that some of the slower-rotating (with $P_{\rm rot}>10$ days) objects produced flares of similar radiated energy, and were apparently single at the resolution of their spectra. \citet{maehara2015} examined the duration of white light flares from these G stars observed by Kepler using short-cadence data, finding flare durations as short as a few minutes.  \citet{namekata2018} added the perspective of solar white light flares, and noted from trends of flare duration and energy that the solar white light flares fall above where the Kepler white light flares would lie in duration, after accounting for the difference in flare energies (their Figure 1). \citet{maehara2015} derived a relationship between flare duration and energy as $\tau_{\rm flare} \propto E_{\rm flare}^{1/3}$, which their data is in good agreement with. They explained the relationship as arising from timescales associated with magnetic reconnection ($\tau_{\rm flare} \sim \tau_{\rm rec} \sim L/(v_{A} M_{A})$), and flare energy arising from extraction of magnetic energy in a spot region ($E_{\rm flare} \sim f E_{\rm mag} \sim f B^{2} L^{3}$), where $L$ is the length scale of the active region, $v_{A}$ is the Alfv\'{e}n speed, $M_{A}$  the reconnection rate, $f$ the efficiency of converting magnetic energy ($E_{\rm mag}$) to flare energy ($E_{\rm flare}$), and $B$ the magnetic field strength in the active region. 
Manipulating these equations gives two scenarios, \\
\begin{equation}
    \tau_{\rm fl} \sim \rho^{1/2} B^{-5/3} E_{\rm fl}^{1/3}
\end{equation}
or \\
\begin{equation}
    \tau_{\rm fl} \sim \rho^{1/2} L^{5/2} E_{\rm fl}^{-1/2} \;\; .
\end{equation}
Fig \ref{fig:namekata} shows the plot of flare duration against flare energy for the NUV flares considered here, in addition to the solar and stellar white-light flares
reported in \citet{maehara2015} and \citet{namekata2018}.

The population of stellar flares detailed here are short-duration, with the bulk of the events lasting less than 5 minutes. A strict comparison of our results for duration vs. energy with those reported in \citet{namekata2018} must first take into account the different bandpasses considered here versus those in the Kepler data.  Using the energy partition calculations in \citet{ostenwolk2015}, particularly their Table 2, we see that assuming a blackbody spectrum dominates the optical wavelength range returns an estimation of $f=E_{\rm opt}/E_{\rm bol}$ of 0.16. As discussed above in \S 3.5, we get a fractional contribution of the radiation in the GALEX NUV bandpass to the total bolometric energy of $p_{\rm bol}=0.13$. The energies calculated in \citet{maehara2012} and subsequent papers are the blackbody energies, which means that they can be further adjusted to be equivalent to our calculated $E_{bol}$ using the ``hot blackbody'' fractionation values given in \citet{ostenwolk2015} Table 2. Comparing with \citet{namekata2018}'s Figure~1,  the durations of NUV stellar flares with GALEX are shorter than those of the solar white-light flares, and the energy range is shifted to larger energies by about two orders of magnitude (Fig. \ref{fig:namekata}). While the energy range is similar to the flares studied in the Kepler bandpass, the durations are much shorter. We note that \citet{Hawley2003}'s study of simultaneous multiwavelength flares on a nearby M dwarf show approximately similar durations for flares seen in the UV as well as V band (their figures 5,7,8,9), bolstering our intercomparison of white-light flares and NUV flares. Interpreting the scaling relations shown in Figure~\ref{fig:namekata}, the relationship between flare energy and duration in the present sample suggests active region length scales of several $10^{9}$-- several 10$^{10}$ cm, or magnetic field strengths in the active region exceeding several hundred Gauss. 

\citet{kowalski2018} found that the time evolution of the NUV continuum is faster than the NUV emission lines, particularly Fe~II. The time evolution of Mg~II  lines tends to follow that of Ca~II \citep{kowalski2018}. \citet{kowalski2018} used a multithread model to describe the time profile and spectra for a hybrid/gradual flare on an M dwarf. In this specification, an average burst spectrum of impulsive heating and impulsive cooling with an extremely energetic beam of nonthermal electrons is used to describe fast evolution on five second timescales. So even these short duration flares may well be composed of multiple emitting loops rather than an elemental structure.

\subsubsection{Light Curve Morphology}

While the diversity of light curve morphology was noted in \S 2, the incidence of occurrence of these different morphologies was not quantified further. Figures~\ref{fig:fredFlares}, ~\ref{fig:symFlares} and ~\ref{fig:multFlares} give examples of FRED-type flares, as well as more symmetric events and those exhibiting multiple sub-peaks. In contrast, studies at optical wavelengths have arrived at a flare template \citep{Davenport2016} to describe the vast majority of flare light curves seen in white light with Kepler. The formation of the NUV is not quite as simple as the white light, and originates over a range of layers in the atmosphere (see discussion above in \S 4.2.1).  Given these two factors, it is not likely that a unique flare template can be derived to describe all NUV stellar flare observations. 

For flare observations with high enough time resolution, additional information can potentially be gleaned from light curve analysis. \citet{solarflareoscillations} pointed out the utility of flare-generated oscillations of coronal loops, which can be used to uniquely constrain the magnetic field strength in coronal closed magnetic structures. This arises by interpreting observed oscillations in intensity during the course of the flare as global standing waves; this enables a connection between the period of the oscillations and the loop's length with the magnetic field strength in the loop. \citet{stellarflareoscillations} reported the first detection of an oscillation in a stellar X-ray flare, noting oscillations at the peak of what was an otherwise flat-topped flare light curve. \citet{doyle2018} reported recently on stellar flare oscillations seen in some GALEX NUV light curves of nearby M dwarfs. The dataset of flaring stars identified here, with access to high time resolution observations of solar-like stars, is ripe for investigating for evidence of oscillations. This endeavor is beyond the scope of the current project.

\section{Conclusion}

This study is remarkable both for the sheer number of UV flares studied, far larger than any other study, as well as for the stellar characteristics, e.g. largely solar-like stars. We analysed $\sim$34,000 light curves from the GALEX space telescope and discovered a population of $\sim$2,000 mostly small, short duration flares that have not been previously cataloged.  We drew our stellar population from the list of stars targeted by both the Kepler and GALEX missions between 2009 and 2012. Because we analyzed a subset of the Kepler target set the majority of our targets were G-type main sequence stars, although the sample ranged in temperature from 3000-11000 K, and in radius from range 0.5-15 R$_{\odot}$. This wavelength  range plus stellar type is a unique probe of stellar flaring and gives insight into the potential conditions on planets orbiting this type of star. A small subset ($\sim$ 15\%) of our flaring stars have rotation periods measured, and the distribution appears to be consistent between flaring and non-flaring stars, indicating that we are seeing flares even on generally inactive stars. This is consistent with the span of quiescent luminosities derived for flaring stars. Because the flares we discovered were small and short, they will be difficult to detect in the Kepler light curves.  Indeed the \citet{Davenport2016} flare survey overlaps very little with our flare detections (only 13 of our flaring stars are also in the Davenport sample), although as discussed in \S 4.1 there are a number of possible explanations for this.

We determined the energy partition for flares observed in the GALEX bandpass and calculated flare energies for all flares on stars for which we could calculate the distance (see Table \ref{tbl:dataRedux} for a break down of our data cuts).  We found that the cumulative energy distribution of our body of flares followed a power law that agrees well with other stellar and solar flare calculations.

In addition to the ancillary information about these stars, we have an independent
measure of magnetic activity in the datasets, namely the quiescent NUV luminosity, in addition to the transient magnetic activity as measured by the flares. We find
correlations of these two magnetic activity signatures, as expected.

The timescales of our sample of flares are very short, again making them a unique population of flares for study.  Figure \ref{fig:namekata} shows timescale vs. energy for several different flare populations, and while our flares are at similar or even smaller timescales than the solar flare population in the figure, our energies are significantly higher, and while our flares are in a similar energy range to the stellar white light flare populations, the timescales are much shorter. However, taking the three stellar flare populations together we do not see a dependence between flare duration and energy, with a spread of three orders of magnitude in duration and a spread of the same magnitude in flare energy. Because of observational limitations we cannot probe stellar flares at the smaller energy ranges we see in solar flare populations, however the flare frequency distribution we calculated on our population is similar to what has been seen on other stars and the Sun (Table \ref{tbl:alphas}) across different wavelength regions. This shows that, at least a high level, these NUV flares are manifestations of the same physical processes. 

By filtering on stellar properties, we find evidence for the most solar-like stars
to have a restricted flare energy range compared to the entire sample, but
which is still elevated with respect to the largest flare energy releases seen on 
the Sun.

A future study will combine the results from this paper with the simultaneous Kepler observations to further constrain the energy partition and examine the predictive ability of optical flare studies to determine the likely NUV flare energy release.

\vspace{5mm}

\acknowledgements
We thank Chase Million, lead gPhoton developer, for all of his help and willingness to answer questions about gPhoton and its best use.
All of the data presented in this paper were obtained from the Mikulski Archive for Space Telescopes (MAST). STScI is operated by the Association of Universities for Research in Astronomy, Inc., under NASA contract NAS5-26555. Support for MAST for non-HST data is provided by the NASA Office of Space Science via grant NNX09AF08G and by other grants and contracts.
This paper includes data collected by the Kepler mission. Funding for the Kepler mission is provided by the NASA Science Mission directorate.
This work has made use of data from the European Space Agency (ESA) mission {\it Gaia} (\url{https://www.cosmos.esa.int/gaia}), processed by the {\it Gaia} Data Processing and Analysis Consortium (DPAC, \url{https://www.cosmos.esa.int/web/gaia/dpac/consortium}). Funding for the DPAC has been provided by national institutions, in particular the institutions participating in the {\it Gaia} Multilateral Agreement.
This work made use of the gaia-kepler.fun crossmatch database created by Megan Bedell.
This research has made use of the SVO Filter Profile Service (http://svo2.cab.inta-csic.es/theory/fps/) supported from the Spanish MINECO through grant AyA2014-55216.\\

Facilities: GALEX (NUV), Kepler

Software: astropy \citep{Astropy}, gPhoton \citep{million2016}

\end{document}